%% file: paper.tex
\documentclass[a4paper,fleqn,usenatbib]{mnras}

\usepackage[T1]{fontenc}
\usepackage{ae,aecompl}

\usepackage{threeparttable}
\usepackage{xspace}

\usepackage{graphicx}	% Including figure files
\usepackage{amsmath}	% Advanced maths commands
\usepackage{amssymb}	% Extra maths symbols

\usepackage{indentfirst}
\usepackage{booktabs}
\usepackage{amsmath,amssymb}
\usepackage{gensymb}
\usepackage{subcaption}
\usepackage[british]{babel}

\newcommand{\AC}{$\alpha$ Cen\xspace}

\title[]{Activity and telluric contamination in HARPS observations of Alpha Centauri B}

\author[Lisogorskyi, M. et al.]{
	Lisogorskyi, M.$^{1}$,
	Jones, H.R.A.$^{1}$,
	and Feng, F.$^{1}$,
\\
$^{1}$Centre for Astrophysics Research, University of Hertfordshire, College Lane, AL10 9AB, Hatfield, UK
}

\date{Accepted 2019 February 24. Received 2019 January 30; in original form 2018 December 11}

\pubyear{2018}

\begin{document}
\label{firstpage}
\pagerange{\pageref{firstpage}--\pageref{lastpage}}
\maketitle

\begin{abstract}

The Alpha Centauri system is the primary target for planet search as it is the closest star system composed of a solar twin \AC A, a K-dwarf \AC B and an M-dwarf Proxima Centauri, which has a confirmed planet in the temperate zone.
\AC A \& B were monitored intensively with the HARPS spectrograph for over 10 years, providing high-precision radial velocity measurements.
In this work we study the available data to better understand the stellar activity and other contaminating signals.
We highlight the importance of telluric contamination and its impact on the radial velocity measurements.
Our suggested procedures lead to discarding about 5\% of HARPS data, providing a dataset with an RMS improved by a factor of 2.
We compile and quantify the behaviour of 345 spectral lines with a wide range of line shapes and sensitivity to activity.

\end{abstract}

\begin{keywords}
exoplanets -- stars: activity -- stars: individual: HD128621
\end{keywords}

%%%%%%%%%%%%%%%%%%%%%%%%%%%%%%%%%%%%%%%%%%%%%%%%%%

%%%%%%%%%%%%%%%%% BODY OF PAPER %%%%%%%%%%%%%%%%%%

%%%%%%%%%%%%%%%%%%%%%%%%
%%%%% INTRODUCTION %%%%%
%%%%%%%%%%%%%%%%%%%%%%%%
\section{Introduction}

The first planetary system was confirmed around the pulsar PSR B1257+12 \citep{1992Natur.355..145W} and the first exoplanet orbiting a solar-type star 51 Pegasi was detected three years later \citep{1995Natur.378..355M}.
Today there are thousands and the number is growing.

The radial velocity (RV) method is currently responsible for detection or confirmation of widest range of exoplanets possible.
This method was proposed years before spectrographs with precision high enough to detect planets became available \citep{1952Obs....72..199S}, with the exception of hot Jupiters.
It is an indirect method that measures the motion of the host star around the barycenter of the star-planet system.
It relies on tiny shifts in narrow absorption features in stellar spectra, which are prone to variations due to stellar sources of noise such as oscillations, granulation, rotating active regions, and magnetic cycles.
The first two can be minimized using an optimised observational strategy \citep{2008MNRAS.386..516O}, while other have much longer time-scales and can produce periodic signals in RV measurements via line profile variations \citep[e.g.,][]{1997ApJ...485..319S, 2001A&A...379..279Q, 2008A&A...489L...9H}.
Others can be identified and corrected for using activity indicators and cross-correlation function asymmetry \citep[e.g.,][]{2001astro.ph..1377S, 2003AA...401.1185L, 2003A&A...403.1077K, 2010A&A...511A..54S}.
These stellar noise sources already have higher RV amplitude than instrumental noise in high-precision spectrographs like HARPS \citep{2003Msngr.114...20M}, so a better understanding of the impact of activity on the spectra and radial velocity measurements is required.
More recent approaches include analysis of differential radial velocities \citep{2017AJ....154..135F}, wavelength-dependent noise via measuring RV from individual spectral lines \citep{2018arXiv180901548D}, and analysis of line shape variations \citep{2017ApJ...846...59D, 2017MNRAS.468L..16T, 2018AJ....156..180W}.

In this work we analyse the available observations of \AC system with HARPS and quantify effect of telluric lines and activity-sensitive lines across the whole spectrum.
Our focus is on \AC B (spectral type K1V) due to a large number of observations which are well spaced in time.
A weak signal was detected in the system before \citep{2012Natur.491..207D}, but it was not confirmed \citep{2013ApJ...770..133H, 2016MNRAS.456L...6R}.
Nevertheless, the star still may have low-mass planets \citep{2018AJ....155...24Z}.

The available data, as well as its quality and rejection criteria, is discussed in Section \ref{sec:data}.
Telluric contamination is estimated and additional observations are rejected in Section \ref{sec:telluric}.
Activity sensitive lines and a list of indices are discussed in Section \ref{sec:activity}.

%%%%%%%%%%%%%%%%%%%%%%%%
%%%%%%%%% DATA %%%%%%%%%
%%%%%%%%%%%%%%%%%%%%%%%%
\section{Data}
\label{sec:data}

The dataset consists of 22559 high resolution ($R$ = 110000) spectra obtained with the HARPS spectrograph between 2005 and 2016, including both \AC A \& B observations.
We use the one-dimensional (so called s1d) spectra and radial velocities produced by the HARPS DRS.

As the stars are very bright, most spectra have signal to noise ratio (SNR) above 100.
The data is naturally divided into observing seasons (one per year) due to observability of the stars.

Despite high SNR on average, some spectra suffer from noise, overexposure, high activity, tellurics or light contamination from the second component of the binary system.
Several methods, as well as visual inspection of the spectra, were employed to minimize the noise.

The analysis is focused on \AC B as it has much better time coverage.
Almost all of \AC A data were taken during five nights in April 2005 and coverage of further seasons is very sparse.
The two components are distinguished using equivalent widths of Na \textsc{d} doublet (see Section \ref{sec:data:visual}).

The data in this work include unbinned observations from seasons considered in \citet{2012Natur.491..207D} with addition of later observations.

High cadence observations in 2013 were excluded from this analysis as most measurements from the spectra (like RV and spectral indices) correlate closely with airmass.
We expect that a combination of two factors are the cause: light contamination from \AC A \citep{2015IJAsB..14..173B} and point-spread function variations over a night \citep{2016MNRAS.459.3551B, 2017MNRAS.469.4268B}.
These observations should be analysed separately.

\subsection{Visual inspection of the spectra}
\label{sec:data:visual}

Na \textsc{d} doublet region of the spectrum was visually inspected and classified.
In total 249 spectra were rejected -- 67 observations removed due to low SNR, 37 due to steep slope of the continuum (change of flux by a factor of 2 within 10 \AA) and 58 due to overexposure.

\begin{figure*}
	\centering
	\begin{subfigure}{.49\textwidth}
		\centering
		\includegraphics[width=\textwidth]{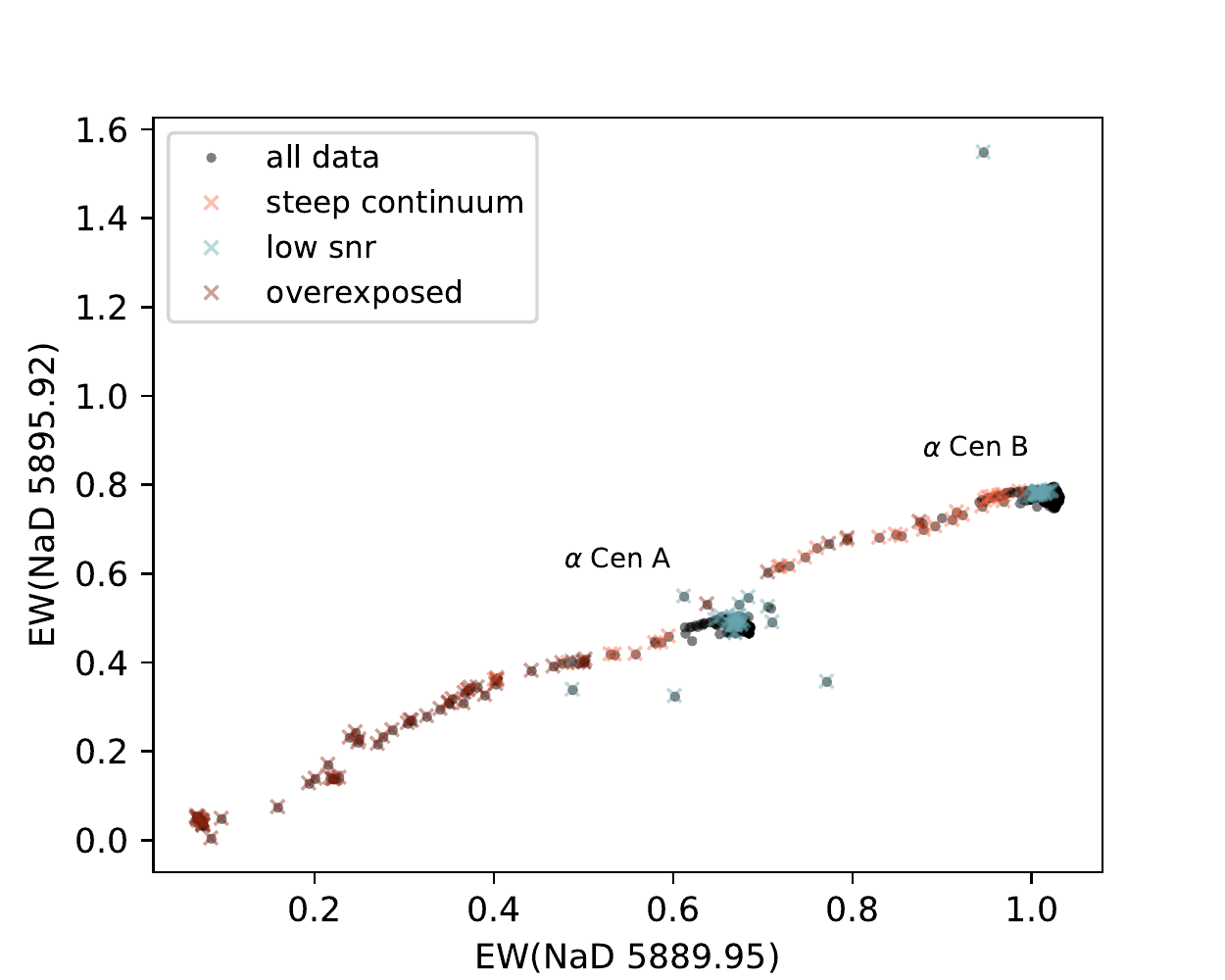}
		\caption{}
		\label{fig:visualnad}
	\end{subfigure}
	\begin{subfigure}{.49\textwidth}
		\centering
		\includegraphics[width=\textwidth]{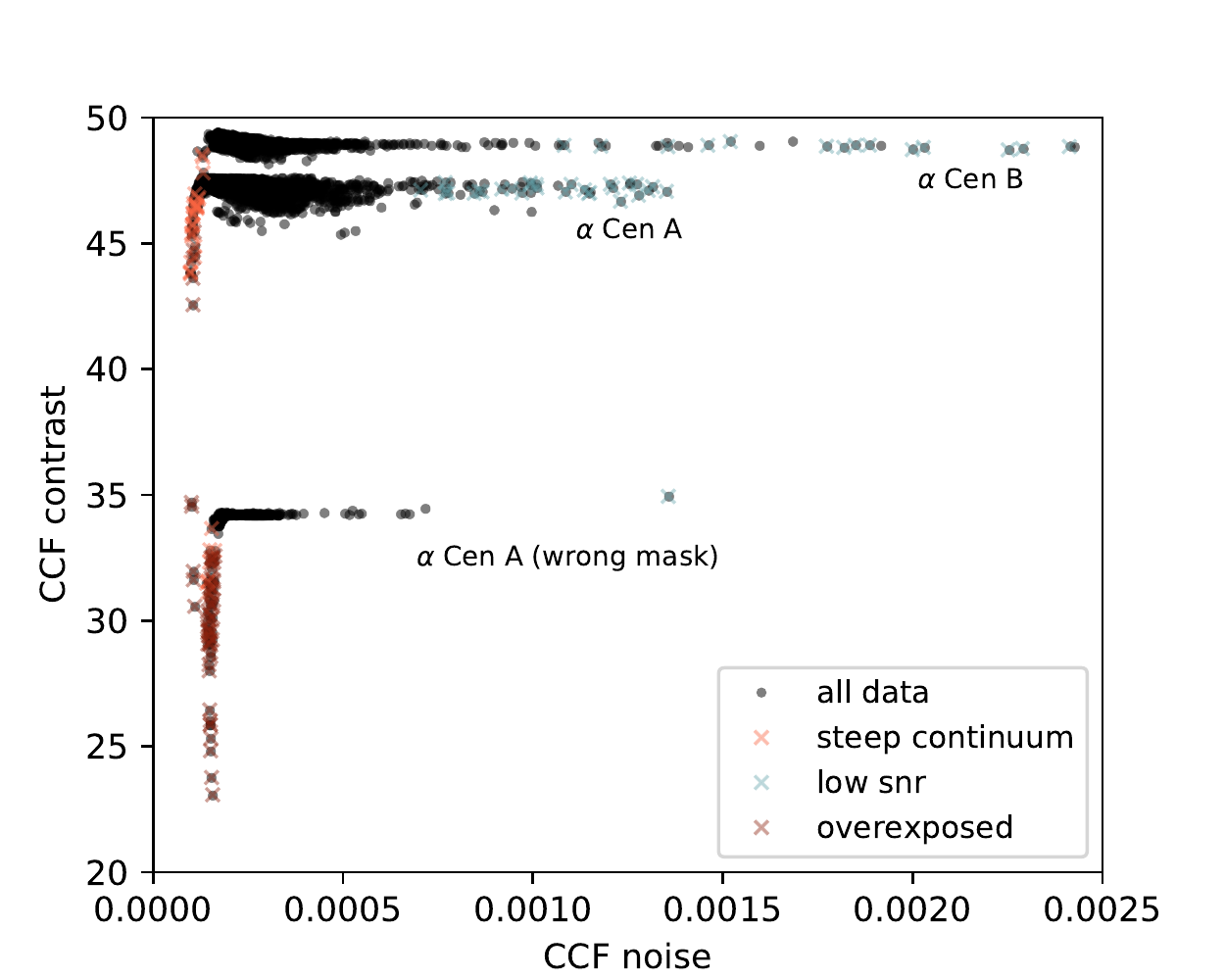}
		\caption{}
		\label{fig:visualccf}
	\end{subfigure}
	\caption{
		Rejected observations based on visual inspection of the spectra.
		Black points are all observations available, `x' show rejected observations due to continuum slope, low SNR and overexposure.
		\textit{Left}: Equivalent widths of Na \textsc{d} doublet.
		\textit{Right}: CCF contrast vs noise.
	}
\label{fig:visual}
\end{figure*}

The rejected spectra are shown in Figure \ref{fig:visual}, showing data distribution in Na \textsc{d} lines equivalent widths and cross-correlation function (CCF) properties.
Most rejected observations are outliers in both plots.
Spectra were identified as ``low SNR'' if the noise had amplitude of the shallow absoption features in that spectral region (most spectra don't show any clearly visible noise).
The limit is to CCF noise of 0.0006 or SNR of 18 in the order 40 (as provided in FITS headers).
Overeposed spectra show ``steppy'' continuum and very narrow sodium lines, as well as steep slope of the continuum, which most likely comes from the wavelength dependence of non-linear regime of the CCD \citep{2012ApJS..200...15A}.

The two components of the binary are well separated on the left plot (Figure \ref{fig:visualnad}) after the bad observations are removed.
All spectra with an equivalent width of Na \textsc{d} line at 5889.95\AA{} below 0.8 were identified as \AC A (4770 spectra), and the ones above - as \AC B (17096 spectra).
The right plot (Figure \ref{fig:visualccf}) shows 3 distinct features--\AC B CCF computed with K5 template mask, \AC A CCF computed with G2 template mask, and \AC A CCF computed with K5 template mask (wrong template mask--about 10\% of \AC A observations affected).

\subsection{CCF properties and spectral indices of \AC B}
\label{sec:data:proxies}

In this section the observing seasons were considered separately and the outliers in the spectral indices and CCF properties were removed based on a visual inspection.
Observations were considered outliers if the values were more than 3$\sigma$ for the particular night or season.
As a result, 205 observations were removed from the sample (see Table \ref{cuts-table}).

Equivalent widths were calculated as 
\begin{equation}
\centering
\label{eq:ew}
EW = (\lambda_1 - \lambda_0) - \sum_{i} \frac{F_i}{F_0} \Delta \lambda,
\end{equation}

where [$\lambda_0$--$\lambda_1$] is the range with the core of the line, $F_i$ is flux in a data point, $F_0$ is continuum level, and $\Delta \lambda$ is resolution of the spectrum (0.01\AA{} for all 1D spectra produced by the HARPS DRS).
The continuum level is calculated by averaging flux in predefined wavelength ranges with few spectral lines on either side of the line core.
Line strengths were calculated from fitting a Gaussian function to the line as $- A/F_0$, where $A$ is the height of the Gaussian and $F_0$ is the base.

The equivalent width of H$\alpha$ was measured at [6561.85--6563.85] \AA{} with continuum measured from [6550.00--6555.00] and [6565.00--6570.00] \AA{}.
Line strength, sigma (from Gaussian fit to the line) and equivalent width of $H_\alpha$ were inspected as both methods quantify changes in the same line and only the outliers in both were removed.
In total 111 observations were discarded ($\sim$0.7 \% of the data).

Equivalent widths of Na \textsc{d} lines were measured at $5889.92$ \AA{} and $5895.92$ \AA{} with a bandpass of 1 \AA{}.
The continuum was estimated from [5800.00--5810.00] and [6080.00--6100.00] \AA{} by averaging the top 10 values from each region \citep[the same spectral ranges as used in][but here the lines are measured separately, as described above]{2011AA...534A..30G}.
The lines were also fitted using four Gaussians - two wide and two narrow components.
Equivalent widths of both lines were investigated as well as gaussian parameters of the wide components (which correlate closely with the narrow ones).
In total 54 outliers were discarded.

Mount-Wilson $S$-index was measured as $S = \alpha \frac{F_H+F_K}{F_R+F_V}$, where $F_H$ is flux measured with triangular filter, centered at 3968.470 \AA{} and with $FWHM = 1.09$,
 $F_K$ is flux measured with triangular filter, centered at 3933.664 \AA{} and with $FWHM = 1.09$,
 $F_R$ is flux measured in [3891, 3911]  \AA{},
 $F_V$ is flux measured in [3991, 4011]  \AA{},
 and $\alpha$ is a constant of proportionality adopted to be 2.4 \citep{1991ApJS...76..383D}.
The value is not corrected for bolometric lumionsity.
$S$-index was inspected using the same procedure and 23 outliers were discarded.

The CCF asymmetry was measured as $V_{span}$ -- a difference in RV between upper and lower parts of a CCF by fitting a Gaussian.
The upper part of the CCF is defined in the range [$-\infty:-1\sigma$][$+1\sigma:+\infty$] and the lower part is defined in the range given by [$-\infty:-3\sigma$][$-1\sigma:+1\sigma$][$+3\sigma:+\infty$] \citep{2011IAUS..273..281B}.
$V_{span}$, CCF FWHM and CCF contrast were inspected and 17 outliers were removed.

\section{Telluric contamination}
\label{sec:telluric}

Telluric contamination can have a large impact on radial velocity measurements via both deep \citep{2014SPIE.9149E..05A} and shallow \citep{2014A&A...568A..35C} absorption features.
We selected a pair of deep water lines to estimate the telluric contamination in RV and provide useful practical limits.

\subsection{Method}

Measuring telluric contamination is quite challenging due to blending of the telluric lines with different atomic lines over the course of a year.
The water line at 6543.9 \AA{} was chosen as a proxy for contamination as it is relatively deep and there are no strong atomic features present in its vicinity.

The centroid of the line was estimated to have a width of 0.1 \AA{}, but as a barycentric correction was already applied to the spectra, the line center moves relative to the stellar spectrum.
Its position was empirically approximated with $0.467 \cdot \mathrm{sin}(\frac{2 \pi(\mathrm{BJD}-2453693.43)}{364.8}) + 6543.92$.
The pseudocontinuum was determined by fitting a second order polynomial to three regions with no significant line absorption - [6538.5--6540.5], [6544.8--6545.2] and [6556.6--6557.2] \AA.

\begin{figure*}
	\centering
	\label{fig:watercorrspec}
	\begin{subfigure}{.49\textwidth}
		\centering
		\includegraphics[width=\textwidth]{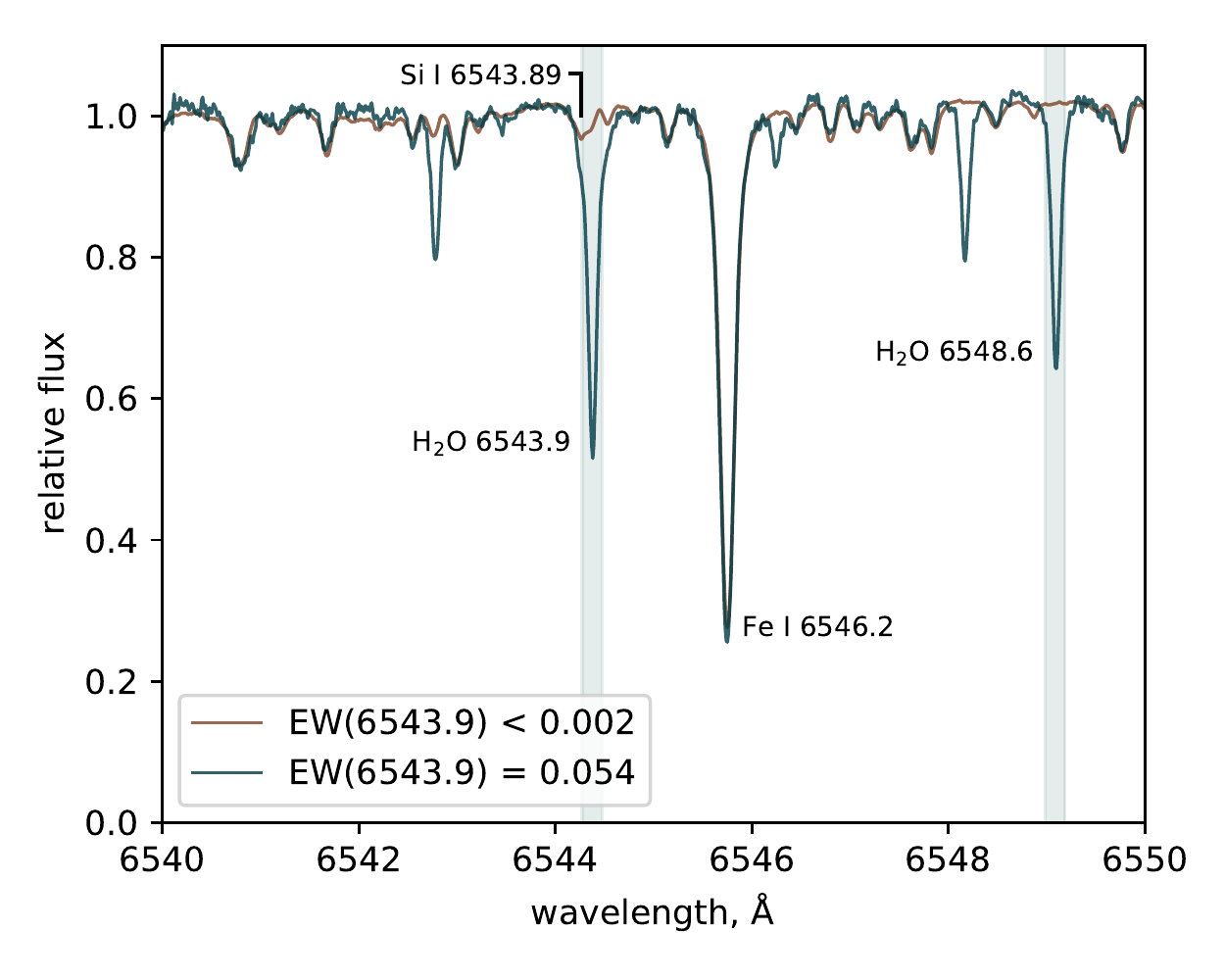}		
		\caption{}
		\label{fig:waterspec}
	\end{subfigure}
	\begin{subfigure}{.49\textwidth}
		\centering
		\includegraphics[width=\textwidth]{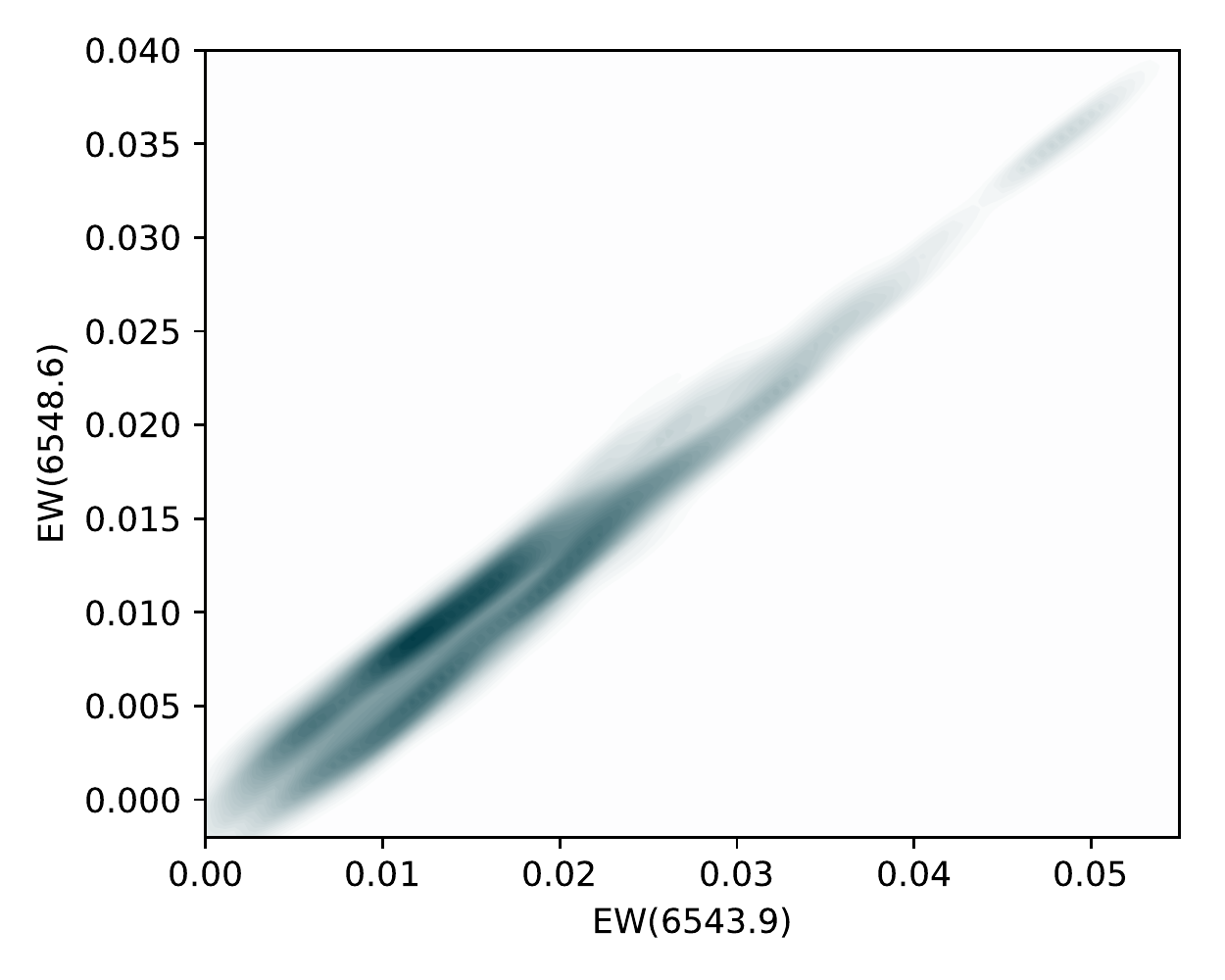}
		\caption{}
		\label{fig:watercorr}
	\end{subfigure}
	\caption{
		Spectral region where the water lines were measured \textit{(a)}.
		The atomic line template spectrum is plotted in red, a high humidity spectrum (\textit{HARPS.2011-03-03T06:52:24.740}, $EW(6543.9) = 0.054$) is in blue.
		The shaded regions show the lines measured, with the width corresponding to the window used to calculate the equivalent width.
		The wavelengths of the marked lines are in rest frame, unlike the measured spectra.
		Comparison plot of the equivalent widths of the two water lines is shown on the right \textit{(b)} as a density plot.
	}
\end{figure*}

To estimate the error, we measured another, neighbouring water line at 6548.617 \AA, which is about twice as weak and blends with slightly stronger atomic lines.

An atomic line template spectrum was produced by averaging RV-corrected spectra with $EW(6543.9\AA) < 0.002$ (all with relative humidity under 10\%).
The telluric lines are virtually non-existent in the selected spectra and so although they have different positions from observation to observation due to the Earth's motion around the Sun they mostly disappear during the averaging.
The template and all the observations were corrected for continua using the pseudocontinuum determination described above and the template was subtracted from the observations for further measurement of the remaining telluric lines.

An example spectrum is shown on Figure \ref{fig:waterspec}.
The atomic template is plotted in red, an example high humidity spectrum is plotted in blue (\textit{HARPS.2011-03-03T06:52:24.740}, $EW(6543.9) = 0.054$).
The shaded regions show the lines measured, with the width corresponding to the window used to calculate the equivalent width.
In the case shown, telluric lines are comparable in depth with narrow absorption featues (i.e. the FeI line in the center of the spectral region), which are crucial for the radial velocity measurements.

The two lines correlate very well with each other (Pearson's $p=0.96$) and the standard deviation is only 0.0014 \AA{} (see Figure \ref{fig:watercorr}).
The duality in the measurement is correlated with the time of year (telluric lines moving relative to the stellar lines), but doesn't produce a large error.
The offset is produced by the water line at 6543.9\AA{} blending with a shallow Si \textsc{i} feature at 6543.89\AA{}.

\begin{figure*}
	\centering
	\label{fig:water_rv}
	\begin{subfigure}{.49\textwidth}
		\centering
		\includegraphics[width=\textwidth]{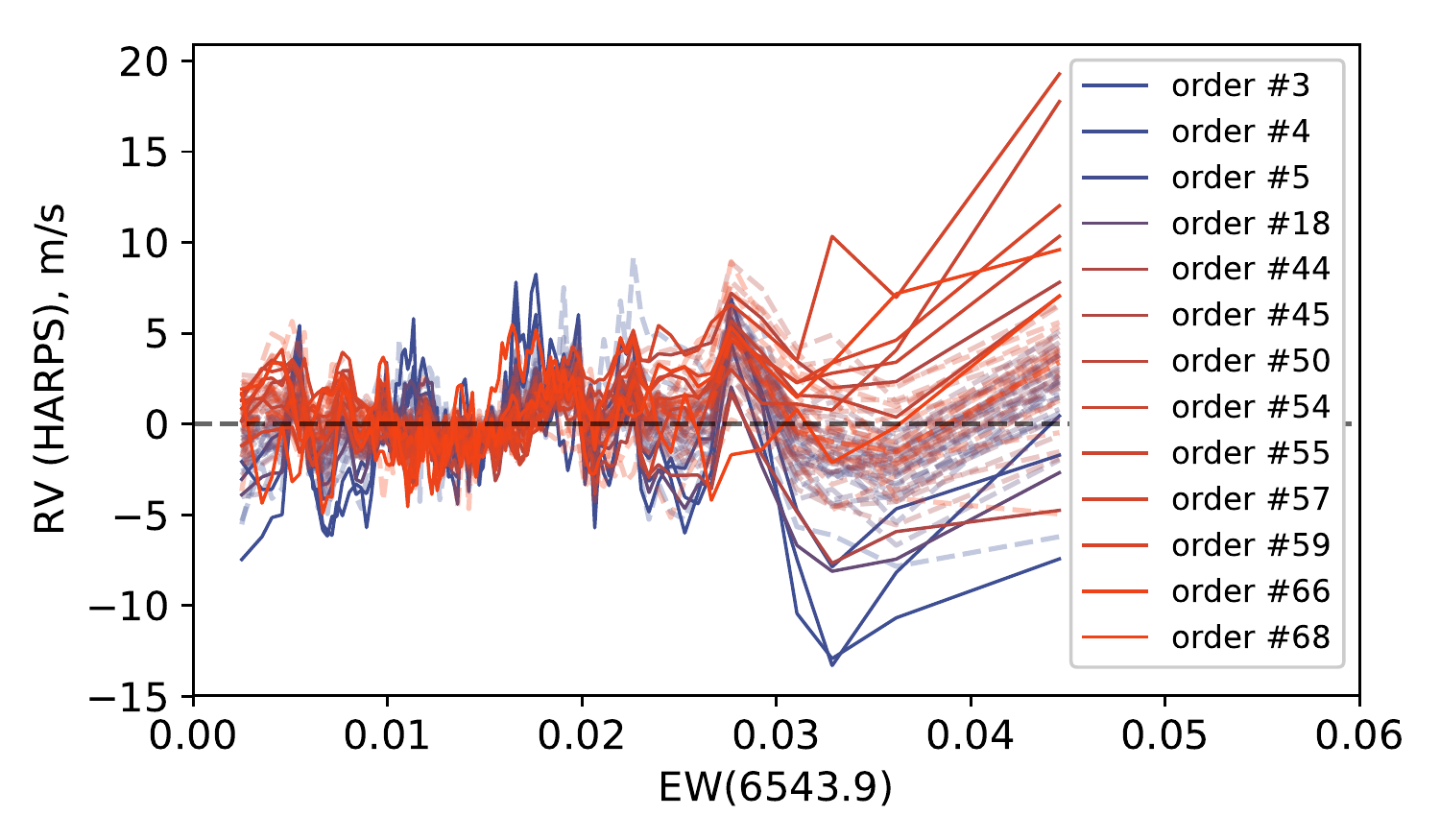}
		\caption{}
		\label{fig:water_rv_a}
	\end{subfigure}
	\begin{subfigure}{.49\textwidth}
		\centering
		\includegraphics[width=\textwidth]{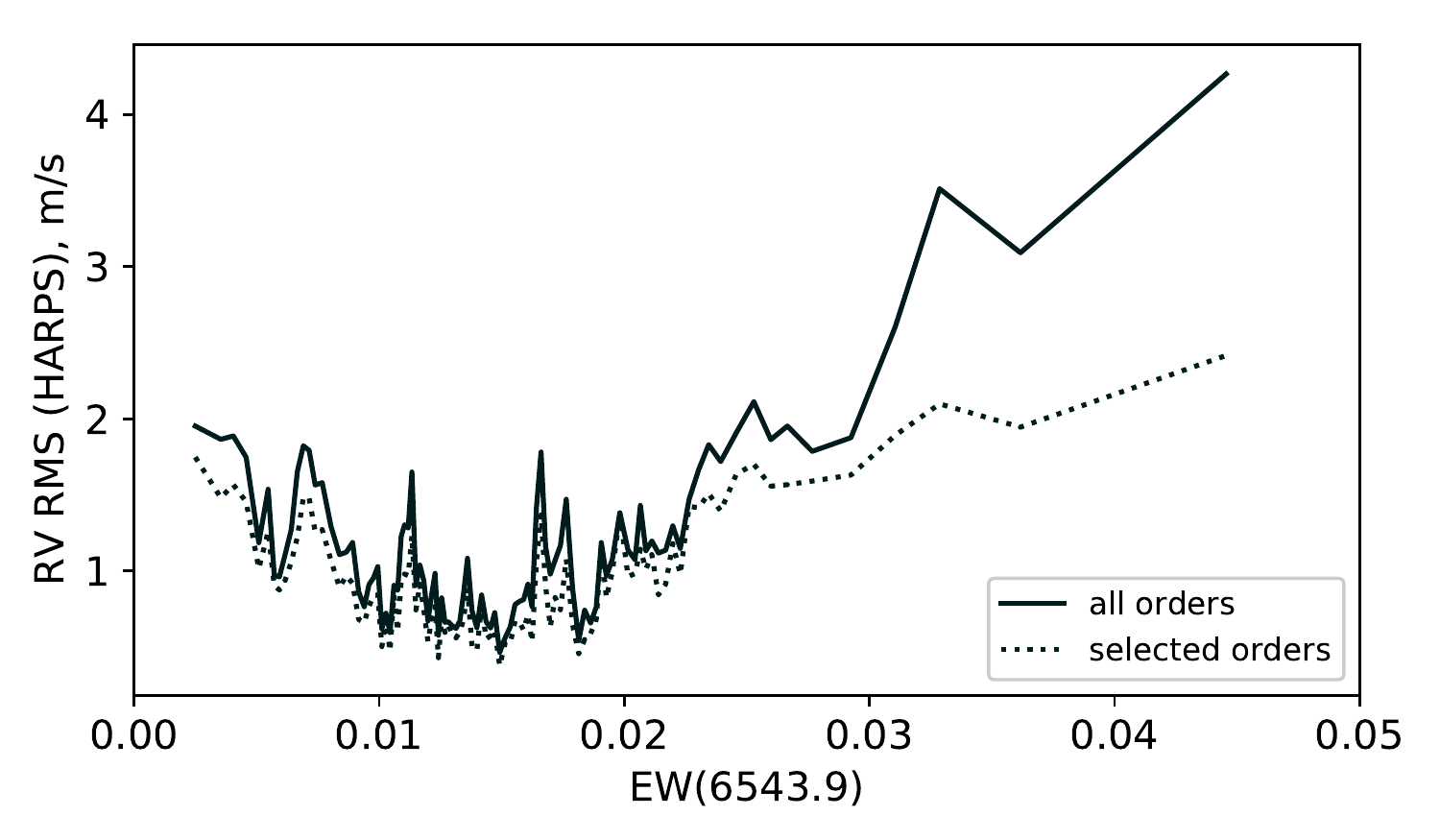}
		\caption{}
		\label{fig:water_rv_b}
	\end{subfigure} 
	\caption{
		Radial velocity measurements versus telluric contamination measured as EW of a water line at 6543.9 \AA{}.
		The data are binned by number of points (167 in each bin).
		Left plot \textit{(a)} shows radial velocity measured from each \'echelle order (color coded appropriately) versus telluric contamination.
		The orders that were found to be impacted by tellurics are plotted as solid lines.
		Right plot \textit{(b)} shows standard deviation as a measure of disagreement between radial velocity measurements from different \'echelle orders.
		Solid line includes all 72 orders, dotted line includes only those identified as less affected by tellurics.
	}
\end{figure*}

\begin{figure*}
	\centering
	\includegraphics[width=0.97\textwidth]{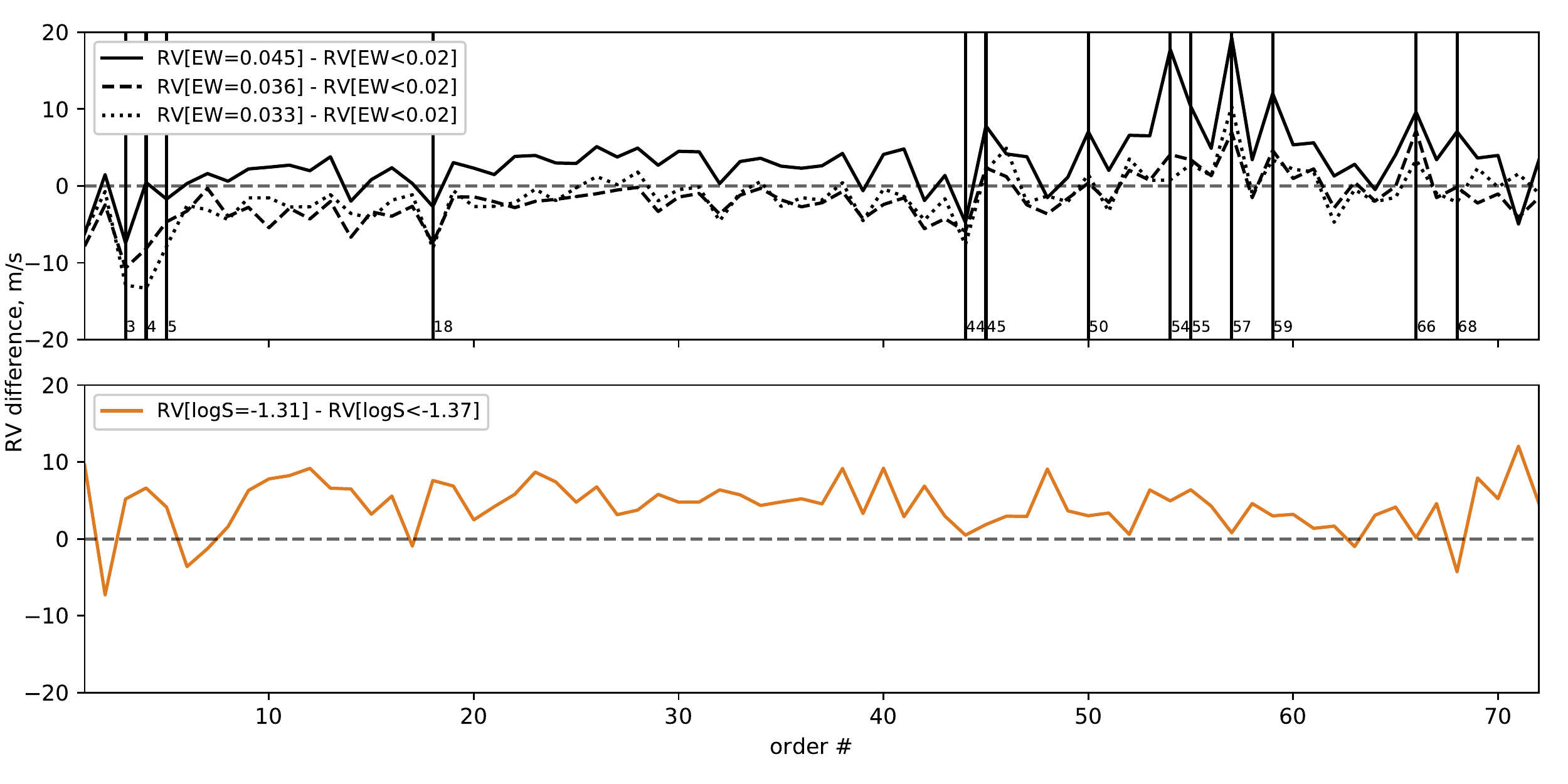}		
	\caption{
		\textit{Top:} the difference between average radial velocity at high telluric contamination (using last 3 bins, see Figure \ref{fig:water_rv_a}) and low telluric contamination (EW < 0.02) for each \'echelle order.
		The orders marked with vertical lines were identified to be affected by telluric contamination the most.
		\textit{Bottom:} similar to the top plot, but the difference between average radial velocity at high activity (log $S$ = -1.31) and low activity (log $S$ < -1.37).
	}
	\label{fig:water_rv_orders}
\end{figure*}

\subsection{Results}
\label{sec:telluric:results}

The impact of telluric contamination on RV measurements was evaluated by subtracting the binary component (third order polynomial fit) from the RV measured from each \'echelle order and plotting it against the EW of the water line (see Figure \ref{fig:water_rv_a}).
The data were binned by the number of points - 100 bins, each containing 167 points.
The RV values on the figure were brought to the same reference point by subtracting mean radial velocity at low humidity ($EW < 0.02$).
At low telluric contamination the orders agree reasonably well, but beyond $EW > 0.03$ some of them start to diverge from the general trend.

The orders that were found to be impacted by tellurics are plotted as solid lines.
This disagreement between RV measurements from different orders was quantified in Figure \ref{fig:water_rv_b} as a standard deviation.
This plot shows two trends - one including all the orders and one with most contaminated orders excluded.

One should keep in mind that the RV data contain a sum of all RV signals in addition to telluric contamination (e.g. activity, rotation, companions etc.).
The threshold of $EW > 0.03$ was chosen as the standard deviation of the RV rapidly increases after this point to twice the value at zero contamination and lowering the threshold does not produce a significant improvement.

The red orders tend to have higher RV scatter with increasing telluric contamination, whereas blue orders tend to do the opposite.
This is quantified by \'echelle order in Figure \ref{fig:water_rv_orders} (top) as a difference of radial velocities in the last 3 bins (high telluric contamination) and low telluric contamination.
The vertical lines mark the most contaminated orders.
This agrees approximately with the finding of \cite{2016ApJ...817L..20W}, who found orders 56--59, 63, 64, 66--68, and 71 too contaminated to be used by their RV extraction method.

Again, as these data contain other signals apart from telluric contamination, we compare to the bottom plot on Figure \ref{fig:water_rv_orders}, which shows a RV effect of activity (as measured by log $S$), using the same binning.
Similarly, orders 3, 4, and 5 show a RV offset, which means that it most likely arises from activity rather than telluric contamination.

This effect can be avoided by excluding the most impacted orders from radial velocity measurements or rejecting all spectra with $EW(6543.9 \AA) > 0.03 \AA$, which is 616 spectra in this dataset.
All cuts with numbers of observations rejected are listed in Table \ref{cuts-table}.

\begin{figure}
	\centering
	\includegraphics[width=0.5\textwidth]{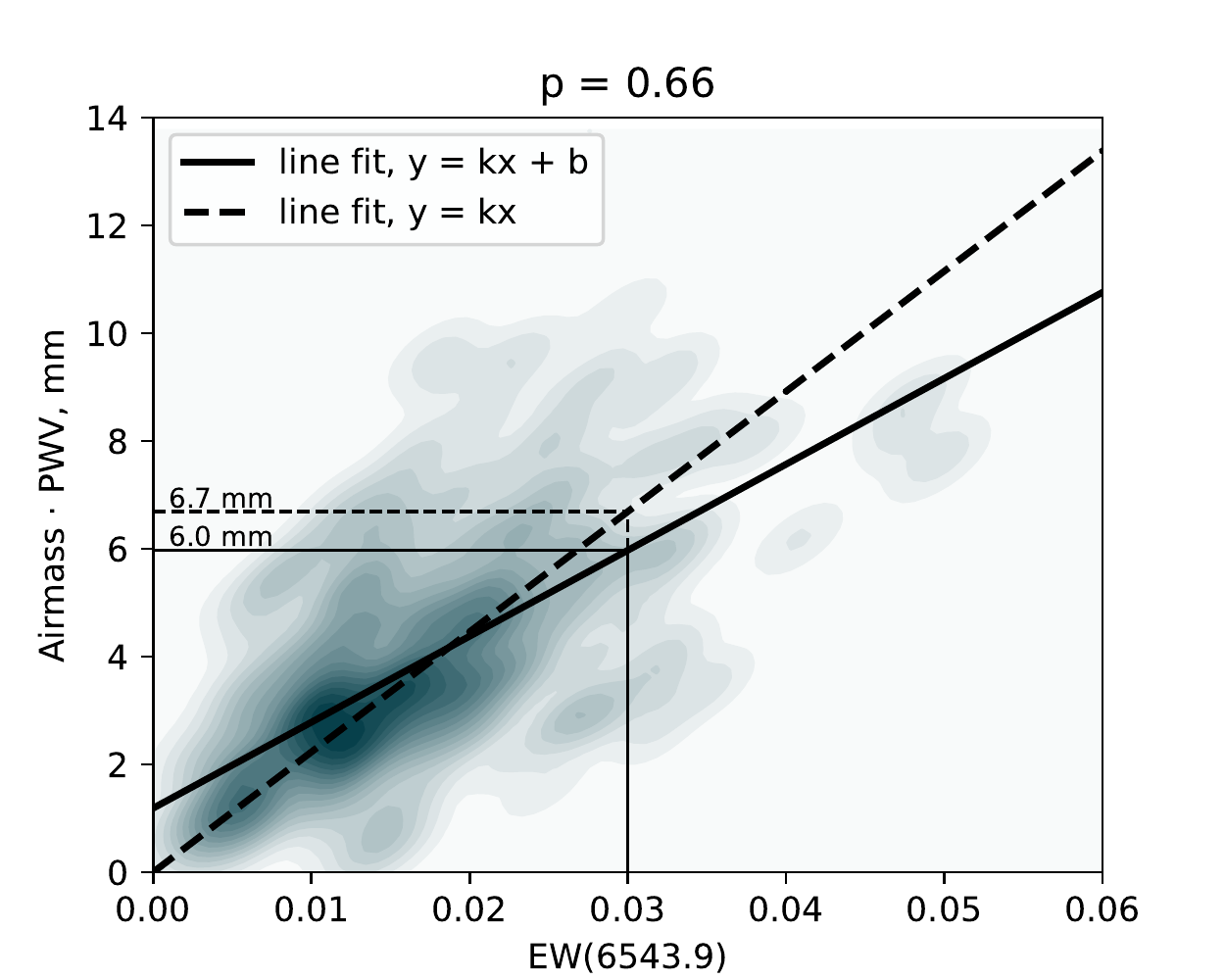}
	\caption{
		Telluric contamination measured as equivalent width of the water line at 6543.9\AA{} versus airmass-corrected precipitable water vapour as a density plot.
		Two line fits are shown in black -- one with an offset (solid line) and one starting at origin (dashed line).
	}
	\label{fig:humid_telluric}
\end{figure}

One can estimate the telluric contamination of the RV measurements using the conditions measured at the site.
The humidity at La Silla is measured at the meteo station using a thin-film polymer humicap at an altitude of 2 meters, as described at the La Silla website\footnote{\url{http://www.ls.eso.org/lasilla/dimm/dimm.html}}.
Comparison of telluric contamination as measured by $EW(6543.9\mbox{\AA})$ and the onsite humidity and temperature measurements recorded is shown on
 Figure \ref{fig:humid_telluric}.
The values of the humidity, temperature, and airmass used were given in FITS headers, apart from 4274 observations in the sample, which have default relative humidity (RH) value of 12\%.
The spectral line measures amount of water vapour along a line of site, so we computed precipitable water vapour (PWV) value using an empirical relationship from \citet{1984SoEn...33..217H}:

\begin{equation}
	PWV = RH (4.7923 + 0.364 T + 0.0055 T^2 +  0.0003 T^3),
\end{equation}
where RH is relative humidity and T is temperature in \degr C.

Here we assume a linear correlation between the sensor readings and the water line depth, so the figure shows two least-squares line fits - one with an offset (solid line) and one starting at origin (dashed line).
The latter represents the assumption of no water line in the spectrum at zero PWV.
The two lines have different slopes, but are still quite close to each other.
Although the two values closely correlate, Pearson's correlation coefficient is only 0.66.
The large scatter around the fits is due to the fact that PWV is calculated using sensors at the observatory site, whereas EW of the water lines measures amount of water along the line of sight.
Presumably, low haze or a thin cloud in a particular direction introduces discrepancy in the measurements.

If we want to remove the most contaminated observations having only the humidity and temperature information provided in the FITS headers, we will need to assume that these fits are good.
This gives us two limits on airmass-corrected PWV -- 6.7 mm and 6.0 mm, as marked on Figure \ref{fig:humid_telluric}.
This way, only 45\% or 58\% of the contaminated data will be removed, respectively.
In addition, 4\% or 8\% extra (uncontaminated) observations are removed if the PWV condition is applied.

Measuring water vapour from the spectra is more robust, as it represents an actual amount of contamination between a target and a telescope.
Removing observations based of the sensor data will result in more contaminated observations left in the dataset due to the scatter.

616 observations of \AC B (3.6\% of the data) in our sample have $EW(6543.9 \AA) > 0.03 \AA$ and were removed from the analysis, providing a cleaner dataset.
To quantify, we subtracted binary component from the RV and computed a root mean square:

\begin{description}
	\item[--] all data: 5.91 m s$^{-1}$,
	\item[--] 2013 excluded: 3.45 m s$^{-1}$,
	\item[--] cleaned dataset: 3.08 m s$^{-1}$.
\end{description}

The RMS value is lower compared to the original dataset as we removed outliers using multiple proxies (CCF properties, spectral indices, tellurics), as opposed to just the RV measurements \citep{2018AJ....155...24Z}.

\begin{table}
	\centering
	\caption{Number of observations removed using different proxies.
	First four rows -- outliers due to light contamination removed using spectral indices and CCF, as described in Section \ref{sec:data:proxies}.
	Last row -- limit on telluric contamination using a water line at 6544 \AA, as described in Section \ref{sec:telluric:results}.
	The final list of observations used is availble as a supplementary online table.
	}
	\label{cuts-table}
	\begin{tabular}{lll}
		\toprule
		Proxy                   & Observations & Observations  \\
								& removed      & removed by    \\
								&              & other proxies \\
		\midrule
		H$\alpha$               & 111          & 0  \\
		Na \textsc{d}           & 73           & 19 \\
		Ca H \& K               & 23           & 0  \\
		$V_{span}$              & 20           & 3  \\
		\rule{0pt}{4ex}EW(6544 \AA) $>$ 0.03   & 616          & 6  \\
		\bottomrule
	\end{tabular}
\end{table}

\section{Activity indicators}
\label{sec:activity}

\AC B is a moderately active K-dwarf with log $R'_{\mathrm{HK}}$ = -4.9 \citep{1996PASP..108..242H, 2012Natur.491..207D}, but still quite a bit of variation is observed.
The most commonly used indicator of chromospheric activity is the $S$-index -- measuring emission in cores of the H and K lines of the Ca II \citep{1978ApJ...226..379W}.
HARPS spectra are high resolution and there are enough data available for \AC B to investigate changes in narrow lines across the whole spectrum.
Three features were identified in a small spectral range [4340\AA, 4480\AA] by \citet{2017MNRAS.468L..16T}.
A strong correlation with Ca II H\&K lines was found for one narrow and two wide features: Fe  \textsc{i}  4375\AA, Fe  \textsc{i}  4383\AA$ $ and Fe  \textsc{i}  4404\AA.
Another 40 were identified across the whole available spectral range by \citet{2018AJ....156..180W} using an automated pipeline.

Here we consider the same season (year 2010) as in \citet{2017MNRAS.468L..16T}, which has clear rotational activity variations and amplitude comparable to long-term activity.
The regular spacing of the observations, sinusoidal variations of the $S$-index with a relatively high amplitude provide both good sampling and baseline for the activity variations.

\subsection{Method}

For each observation from 2010 the spectra were corrected for RV (as measured with the HARPS pipeline) and divided by a low activity template.
The low activity template was computed by median stacking the 10 spectra with lowest values of $S$-index.
The resulting general shape is well approximated with a third order polynomial, but due to an imperfect blaze correction some spectra show continuum variations.
To correct for this, the spectra were median binned by 10 \AA{} and intepolated using cubic splines to identify the pseudocontinuum.
The resulting relative spectra were overplotted and coloured based on $S$-index value for easy identification of active lines.
We visually inspected the whole HARPS spectrum range for different line shape variations.
A line was added to the list if its relative flux variation was above 2\% (the relative spectra have about 1\% noise level).
The species were determined by comparing to the spectral atlas \textsc{spectroweb}.

Equivalent widths of the lines (and their variations) were measured and asymmetries of the lines were quantified by subtracting mean flux in the right and left wings of the line.
Equivalent widths were measured the same way as in the previous section (see eq. \ref{eq:ew}).
The line core was selected manually to contain all the flux variations, and the pseudocontinua were chosen to be non-varying parts of the spectra.
Right and left wings here are parts of a line on either side of a measured center with the same width (half width of the line core window).

In addition, we computed Pearson's correlation coefficients for both measured values compared to log$S$.
The code used in this work to measure line variations is available on \textsc{github}\footnote{\url{https://github.com/timberhill/slice}}.

\subsection{Results}

The full list of 345 activity sensitive lines are given in the appendix \ref{tab:A1}.
These include the 40 lines compiled by \citet{2018AJ....156..180W} and maybe comparable to the 489 lines found by \citet{2018arXiv180901548D}.
Visual inspection of the relative spectra produced much better results in the blue part of the spectrum which is less hampered by tellurics despite the higher noise and relative lack of continuum to rely on. 

Ca H \& K lines were used as a test case for correlation and an estimate of measurement discrepancy between Mount-Wilson $S$-index and equivalent width from a relative spectrum.
The correlation here has very small scatter and correlation coefficient close to 1, which is expected.
Most lines correlate with $S$-index in either flux or asymmetry.
Some are definitely sensitive to activity, but don't correlate with Ca H \& K lines, which could mean the variations are produced via different processes.
In addition, several spectral lines identified in \citet{2018AJ....156..180W} (e.g. Fe \textsc{i} 4602.95\AA) do not show correlation between the equivalent width and $S$-index.
This is most likely because correlations in \citet{2018AJ....156..180W} are obtained between $S$-index and core depth, central width, and RV, which measure different line properties that vary independently from each other.

\begin{figure*}
	\centering
	\begin{subfigure}{.49\textwidth}
		\centering
		\includegraphics[width=\textwidth]{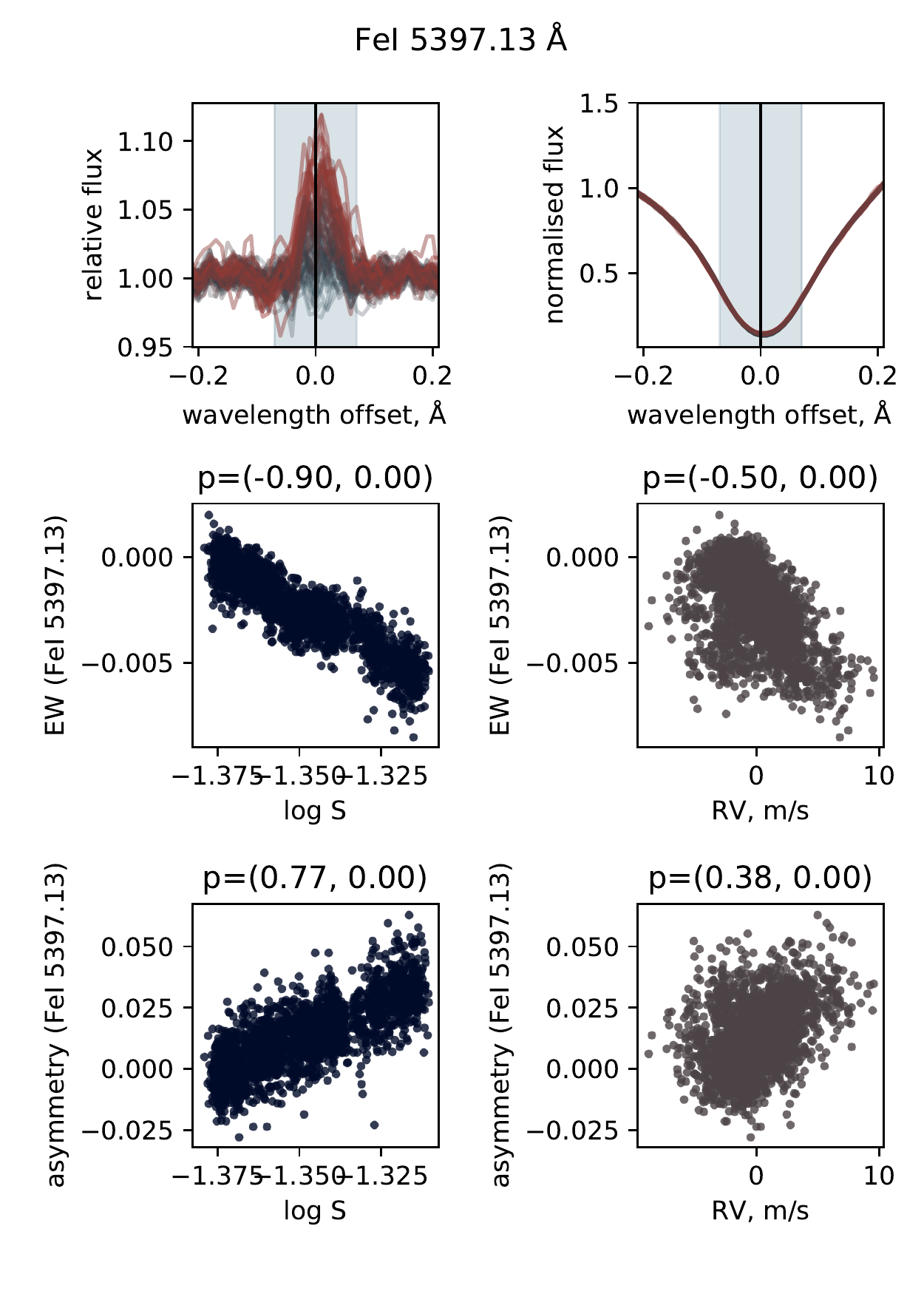}
		\caption{}
		\label{fig:active_flux}
	\end{subfigure}
	\begin{subfigure}{.49\textwidth}
		\centering
		\includegraphics[width=\textwidth]{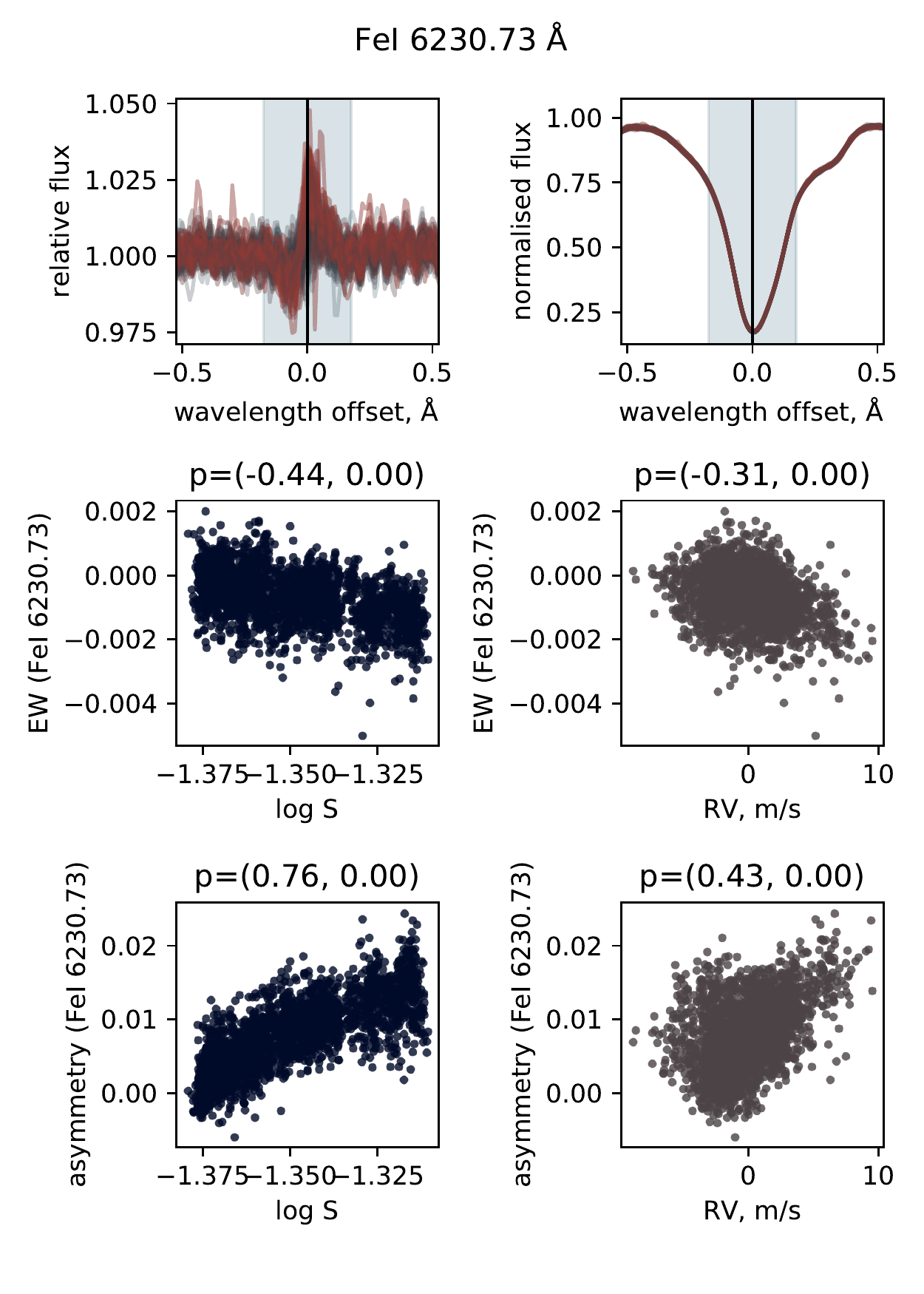}
		\caption{}
		\label{fig:active_asym}
	\end{subfigure} 
	\caption{
		Two examples of the activity-sensitive lines, substantially changing in flux \textit{(a)} and shape \textit{(b)}.
		The plots show the relative spectrum (upper left, red indicates high activity spectrum, black--low activity), normalised spectrum (upper right, same colour scheme), log$S$ vs EW of the line (middle left), log$S$ vs Asymmetry of the line (bottom left), RV vs EW of the line (middle right), RV vs Asymmetry of the line (bottom right).
		}
	\label{fig:relspec_examples}
\end{figure*}

All the lines listed vary in time in different ways -- core flux, position or shape.
Inspecting the relative spectra reveals much more information than just the equivalent width of a line.
Figure \ref{fig:relspec_examples} shows two examples of lines changing in flux and in shape:

\begin{enumerate}

\item Fe \textsc{i} 5397.13\AA{} line (Figure \ref{fig:active_flux}) changes substantially with activity and seems rather symmetrical.
\item Fe \textsc{i} 6230.73\AA{} line (Figure \ref{fig:active_asym}) varies much less in total flux (only about 3\%), but rather in the difference of a mean flux in the red and blue wings.

\end{enumerate}

All lines were compared to the RV measurements from HARPS DRS, but none show a strong correlation, apparently different lines are affected by the activity in different ways and therefore produce different RV shifts.
Radial velocity measured from separate lines would show a strong correlation with line variations \citep{2018arXiv180901548D}.
Even neighbouring lines like Fe \textsc{i} 5225.53 and Fe \textsc{i} 5227.19 show correlation and anti-correlation in flux, respectively.
The same is happening to Cr \textsc{i} 4616.13 and Cr \textsc{i} 4621.93 in both flux and asymmetry variations.

Another type of features is blended lines, of which there are at least 25 in the list.
Two species were specified for one line if two strong lines blended and the resulting feature had only one peak.
If the two lines change in flux in a different way, the resulting blended feature will change shape and mock radial velocity shifts.
We suspect that a substantial amount of variations are caused by blending of strong lines with weaker features that also vary, but are not measurable in a reliable way, even with these high-SNR data.

\section{Discussion}

We analysed HARPS observations of \AC B and assessed the quality of the data.
A number of observations were rejected due to bad quality spectra using independent proxies, providing a cleaner dataset.

Telluric contamination was estimated using a pair of water lines in the spectrum.
Those lines proved to be relatively good indicators as they don't blend with strong atomic features in the stellar spectrum over the course of a year.
We put a limit on contamination using radial velocity measurements of separate orders and identify those that are most affected.
Only a small fraction of observations was too affected to be used.
Removing the affected orders from RV measurements reduced disagreement between orders from about 4 m s$^{-1}$ to 2 m s$^{-1}$.

Activity in FGK stars is commonly estimated from only two Ca H \& K lines.
In this work we present 345 spectral lines sensitive to activity.
The number is much higher than in previous works for a variety of reasons including our focus on high resolution, high signal-to-noise ratio spectra of a single star.
We have visually inspected the selected lines to ensure robust identification of weaker lines and a variety of line shapes.

The line list includes lines throughout the spectral range considered including multiple lines in most spectral orders.
Visual inspection of the relative spectra enables identification of a wide variety of different shape variations of the spectral lines that ultimately affect the radial velocity measurements.
Not all lines in the list correlate with Ca H \& K lines, as measured by the equivalent width variations and asymmetry, but all change in flux or shape (as expected due to the method used).
As the lines originate in different pressure and temperature environments, they vary in different ways that affect radial velocity measurements differently - core flux, wings, width, asymmetries, blended lines etc.
Even neighbouring spectral lines are affected differently, so measuring radial velocity from individual lines \citep{2018arXiv180901548D} of small wavelength ranges \citep{1996PASP..108..500B} will produce much better results.
The list in the Appendix \ref{tab:A1} samples the whole HARPS spectral range and can be treated as a list of the most sensitive lines.

\section*{Acknowledgements}

Based on observations collected at the European Organisation for Astronomical Research in the Southern Hemisphere under ESO programmes 076.C-0878(B), 074.C-0012(B), 072.C-0513(D), 082.C-0315(A), 183.C-0972(A), 083.C-1001(A), 084.C-0229(A), 085.C-0318(A), 088.C-0011(A), 091.C-0844(A), 086.C-0230(A), 087.C-0990(A) and 089.C-0050(A).

This research made use of \textsc{numpy} \citep{van2011numpy}, \textsc{astropy}, a community-developed core Python package for Astronomy \citep{2013A&A...558A..33A}, \textsc{scipy} \citep{jones_scipy_2001}, \textsc{scikit-learn} \citep{mckinney}, \textsc{matplotlib}, a Python library for publication quality graphics \citep{Hunter:2007}.
Spectral features were identified using \textsc{spectroweb}\footnote{\url{http://spectra.freeshell.org}}, interactive digital spectral atlases of bright stars \citep{2006IAUJD...4E..22L, 2008JPhCS.130a2015L, 2011CaJPh..89..395L}.

ML is supported by a University of Hertfordshire PhD studentship.
HJ and FF acknowledge support from the UK Science and Technology Facilities Council [ST/M001008/1].

%%%%%%%%%%%%%%%%%%%%%%%%%%%%%%%%%%%%%%%%%%%%%%%%%%
%%%%%%%%%%%%%%%%%%%% REFERENCES %%%%%%%%%%%%%%%%%%`


\bibliographystyle{mnras}
\bibliography{paper}

%%%%%%%%%%%%%%%%%%%%%%%%%%%%%%%%%%%%%%%%%%%%%%%%%%
%%%%%%%%%%%%%%%%% APPENDICES %%%%%%%%%%%%%%%%%%%%%

\appendix\section{Activity sensitive lines identified in this work}\label{tab:A1}
\include{appendix}

%%%%%%%%%%%%%%%%%%%%%%%%%%%%%%%%%%%%%%%%%%%%%%%%%%

% Don't change these lines
\bsp	% typesetting comment
\label{lastpage}
\end{document}

%% file: appendix.tex
\begin{table*}
	\begin{threeparttable}
		\centering
		\caption{The activity sensitive lines presented below are also available as a supplementary online table.
The table includes measured center, species, equivalent width of the line (EW, measured from a low activity template),
change in equivalent width ($\Delta$EW, measured from relative spectra), 
Pearson's correlation coefficient and 2-tailed $p$-value for log$S$ vs $\Delta$EW ($\Delta$EW PCC), 
asymmetry of the line -- difference between mean flux in the red and blue wings.
Pearson's correlation coefficient and 2-tailed $p$-value for log$S$ vs Asymmetry (Asym. PCC).
		}
		\label{tab:actlines}
		\begin{tabular}{llp{2cm}p{2cm}p{3cm}p{2cm}p{2cm}}
			\toprule
			Center	& Species	& EW	& $\Delta$EW$^1$	& $\Delta$EW PCC$^2$	& Asym.$^3$	& Asym. PCC$^4$	\\
			\midrule
3824.45	&  Fe \textsc{i}	&  0.1170	&  -0.0202	&  (-0.38, 0.00)	&  0.0132	&  (0.12, 0.00)	\\ 
3849.01	&  La \textsc{ii}	&  0.0931	&  -0.0177	&  (-0.51, 0.00)	&  0.0066	&  (0.12, 0.00)	\\ 
3856.38	&  Fe \textsc{i}	&  0.1264	&  -0.0150	&  (-0.28, 0.00)	&  -0.0478	&  (0.07, 0.00)	\\ 
3859.92	&  Fe \textsc{i}	&  0.1689	&  -0.0253	&  (-0.46, 0.00)	&  0.0026	&  (0.07, 0.00)	\\ 
3878.02	&  Fe \textsc{i}	&  0.1123	&  -0.0151	&  (-0.35, 0.00)	&  0.0018	&  (0.21, 0.00)	\\ 
3878.58	&  Fe \textsc{i}	&  0.1122	&  -0.0313	&  (-0.61, 0.00)	&  0.1114	&  (0.44, 0.00)	\\ 
3886.28	&  Fe \textsc{i}	&  0.1381	&  -0.0118	&  (-0.38, 0.00)	&  0.0409	&  (0.18, 0.00)	\\ 
3895.66	&  Fe \textsc{i}	&  0.1034	&  -0.0088	&  (-0.38, 0.00)	&  0.0736	&  (0.21, 0.00)	\\ 
3899.71	&  Fe \textsc{i}	&  0.0655	&  -0.0079	&  (-0.44, 0.00)	&  0.0301	&  (0.15, 0.00)	\\ 
3905.53	&  Si \textsc{i}	&  0.1317	&  -0.0079	&  (-0.40, 0.00)	&  0.0154	&  (0.07, 0.00)	\\ 
3907.94	&  Fe \textsc{i}	&  0.0614	&  -0.0049	&  (-0.33, 0.00)	&  0.0219	&  (0.20, 0.00)	\\ 
3908.76	&  Cr \textsc{i}	&  0.0957	&  0.0052	&  (0.31, 0.00)	&  0.0035	&  (0.10, 0.00)	\\ 
3914.33	&  Ti \textsc{i} + V \textsc{ii}	&  0.1408	&  -0.0012	&  (-0.24, 0.00)	&  -0.0008	&  (0.20, 0.00)	\\ 
3917.19	&  Fe \textsc{i}	&  0.1379	&  -0.0039	&  (-0.33, 0.00)	&  0.0155	&  (-0.01, 0.64)	\\ 
3920.26	&  Fe \textsc{i}	&  0.0658	&  -0.0030	&  (-0.48, 0.00)	&  0.0157	&  (0.23, 0.00)	\\ 
3922.92	&  Fe \textsc{i}	&  0.1227	&  -0.0108	&  (-0.67, 0.00)	&  0.0065	&  (0.19, 0.00)	\\ 
3927.93	&  Fe \textsc{i}	&  0.1218	&  -0.0124	&  (-0.50, 0.00)	&  0.0216	&  (0.29, 0.00)	\\ 
3930.30	&  Fe \textsc{i}	&  0.1294	&  -0.0126	&  (-0.55, 0.00)	&  0.0254	&  (0.20, 0.00)	\\ 
3933.68	&  Ca \textsc{k}	&  0.4170	&  -0.3205	&  (-0.97, 0.00)	&  -0.0113	&  (-0.18, 0.00)	\\ 
3944.01	&  Al \textsc{i}	&  0.9292	&  -0.0255	&  (-0.67, 0.00)	&  0.0010	&  (0.21, 0.00)	\\ 
3961.53	&  Al \textsc{i}	&  0.9980	&  -0.0391	&  (-0.63, 0.00)	&  0.0023	&  (0.00, 0.92)	\\ 
3968.47	&  Ca \textsc{h}	&  0.3971	&  -0.2592	&  (-0.97, 0.00)	&  0.0323	&  (0.64, 0.00)	\\ 
4020.90	&  Co \textsc{i}	&  0.1019	&  -0.0024	&  (-0.33, 0.00)	&  0.0021	&  (0.15, 0.00)	\\ 
4032.63	&  Fe \textsc{i}	&  0.0611	&  -0.0028	&  (-0.34, 0.00)	&  0.0254	&  (0.39, 0.00)	\\ 
4045.82	&  Fe \textsc{i}	&  1.5571	&  -0.0300	&  (-0.79, 0.00)	&  0.0026	&  (0.25, 0.00)	\\ 
4048.75	&  Mn \textsc{i}	&  0.1018	&  -0.0015	&  (-0.21, 0.00)	&  0.0120	&  (0.14, 0.00)	\\ 
4055.55	&  Mn \textsc{i}	&  0.0874	&  -0.0036	&  (-0.47, 0.00)	&  0.0093	&  (0.21, 0.00)	\\ 
4058.93	&  Mn \textsc{i} + Ca \textsc{i}	&  0.0688	&  -0.0011	&  (-0.34, 0.00)	&  0.0168	&  (0.12, 0.00)	\\ 
4061.73	&  Mn \textsc{i}	&  0.0937	&  -0.0028	&  (-0.59, 0.00)	&  -0.0058	&  (-0.07, 0.00)	\\ 
4063.60	&  Fe \textsc{i}	&  1.2458	&  -0.0350	&  (-0.78, 0.00)	&  0.0033	&  (0.08, 0.00)	\\ 
4071.74	&  Fe \textsc{i}	&  1.1328	&  -0.0190	&  (-0.80, 0.00)	&  -0.0006	&  (0.14, 0.00)	\\ 
4077.72	&  Sr \textsc{ii}	&  0.1102	&  -0.0014	&  (-0.04, 0.08)	&  0.0145	&  (0.21, 0.00)	\\ 
4092.68	&  V \textsc{i}	&  0.1091	&  -0.0025	&  (-0.60, 0.00)	&  -0.0226	&  (-0.08, 0.00)	\\ 
4099.79	&  V \textsc{i}	&  0.0772	&  -0.0014	&  (-0.43, 0.00)	&  0.0121	&  (0.28, 0.00)	\\ 
4100.74	&  Fe \textsc{i}	&  0.0702	&  -0.0006	&  (-0.20, 0.00)	&  0.0283	&  (0.40, 0.00)	\\ 
4102.94	&  Si \textsc{i}	&  0.0765	&  -0.0022	&  (-0.63, 0.00)	&  0.0126	&  (0.30, 0.00)	\\ 
4109.80	&  Fe \textsc{i} + V \textsc{i}	&  0.1047	&  -0.0018	&  (-0.53, 0.00)	&  -0.0106	&  (-0.24, 0.00)	\\ 
4110.54	&  Co \textsc{i}	&  0.0949	&  -0.0021	&  (-0.57, 0.00)	&  -0.0009	&  (0.08, 0.00)	\\ 
4111.78	&  V \textsc{i}	&  0.0786	&  -0.0022	&  (-0.63, 0.00)	&  0.0047	&  (0.24, 0.00)	\\ 
4115.18	&  V \textsc{i}	&  0.0837	&  -0.0019	&  (-0.53, 0.00)	&  0.0153	&  (0.26, 0.00)	\\ 
4116.48	&  V \textsc{i}	&  0.0573	&  -0.0019	&  (-0.67, 0.00)	&  0.0141	&  (0.27, 0.00)	\\ 
4116.55	&  V \textsc{i}	&  0.0566	&  -0.0022	&  (-0.79, 0.00)	&  0.0037	&  (-0.26, 0.00)	\\ 
4121.32	&  Co \textsc{i}	&  0.0891	&  -0.0032	&  (-0.40, 0.00)	&  0.0300	&  (0.38, 0.00)	\\ 
4121.81	&  Fe \textsc{i}	&  0.0936	&  0.0025	&  (0.63, 0.00)	&  -0.0018	&  (-0.12, 0.00)	\\ 
4132.06	&  Fe \textsc{i}	&  0.3144	&  -0.0072	&  (-0.73, 0.00)	&  0.0015	&  (-0.06, 0.01)	\\ 
4143.87	&  Fe \textsc{i}	&  0.0908	&  -0.0031	&  (-0.32, 0.00)	&  0.0248	&  (0.34, 0.00)	\\ 
4147.67	&  Fe \textsc{i}	&  0.0619	&  -0.0026	&  (-0.36, 0.00)	&  0.0352	&  (0.41, 0.00)	\\ 
4174.92	&  Fe \textsc{i}	&  0.0782	&  -0.0035	&  (-0.55, 0.00)	&  -0.0069	&  (-0.13, 0.00)	\\ 
4177.60	&  Fe \textsc{i}	&  0.0646	&  -0.0015	&  (-0.44, 0.00)	&  0.0070	&  (0.03, 0.14)	\\ 
4180.83	&  V \textsc{ii}	&  0.1390	&  -0.0048	&  (-0.80, 0.00)	&  0.0088	&  (0.40, 0.00)	\\ 
4187.04	&  Fe \textsc{i}	&  0.0803	&  -0.0002	&  (0.06, 0.01)	&  0.0256	&  (0.38, 0.00)	\\ 
4190.71	&  Co \textsc{i}	&  0.1093	&  -0.0034	&  (-0.81, 0.00)	&  0.0096	&  (0.38, 0.00)	\\ 
4206.70	&  Fe \textsc{i}	&  0.0628	&  -0.0038	&  (-0.77, 0.00)	&  0.0198	&  (0.20, 0.00)	\\ 
4216.19	&  Fe \textsc{i}	&  0.0715	&  -0.0062	&  (-0.87, 0.00)	&  0.0186	&  (0.34, 0.00)	\\ 
4226.73	&  Ca \textsc{i}	&  2.0420	&  -0.0550	&  (-0.93, 0.00)	&  0.0024	&  (0.21, 0.00)	\\ 
4246.83	&  Sc \textsc{ii}	&  0.0713	&  -0.0009	&  (-0.03, 0.23)	&  0.0147	&  (0.49, 0.00)	\\ 
4250.12	&  Fe \textsc{i}	&  0.0764	&  0.0002	&  (-0.03, 0.16)	&  0.0422	&  (0.40, 0.00)	\\ 
4250.79	&  Fe \textsc{i}	&  0.0870	&  -0.0018	&  (-0.35, 0.00)	&  0.0360	&  (0.42, 0.00)	\\ 
4252.30	&  Co \textsc{i}	&  0.0744	&  -0.0031	&  (-0.80, 0.00)	&  0.0101	&  (0.23, 0.00)	\\ 
4254.34	&  Cr \textsc{i}	&  0.5104	&  -0.0225	&  (-0.80, 0.00)	&  0.0127	&  (0.37, 0.00)	\\ 

			\bottomrule
		\end{tabular}
	\end{threeparttable}
\end{table*}
\begin{table*}
	\begin{threeparttable}
		\centering
		\contcaption{The activity sensitive lines presented below are also available as a supplementary online table.
The table includes measured center, species, equivalent width of the line (EW, measured from a low activity template),
change in equivalent width ($\Delta$EW, measured from relative spectra), 
Pearson's correlation coefficient and 2-tailed $p$-value for log$S$ vs $\Delta$EW ($\Delta$EW PCC), 
asymmetry of the line -- difference between mean flux in the red and blue wings.
Pearson's correlation coefficient and 2-tailed $p$-value for log$S$ vs Asymmetry (Asym. PCC).
		}
		\label{tab:actlines}
		\begin{tabular}{llp{2cm}p{2cm}p{3cm}p{2cm}p{2cm}}
			\toprule
			Center	& Species	& EW	& $\Delta$EW$^1$	& $\Delta$EW PCC$^2$	& Asym.$^3$	& Asym. PCC$^4$	\\
			\midrule
4258.32	&  Fe \textsc{i}	&  0.0731	&  -0.0016	&  (-0.36, 0.00)	&  0.0192	&  (0.42, 0.00)	\\ 
4281.37	&  Ti \textsc{i}	&  0.0426	&  -0.0011	&  (-0.58, 0.00)	&  -0.0190	&  (-0.20, 0.00)	\\ 
4291.47	&  Fe \textsc{i}	&  0.0680	&  -0.0021	&  (-0.65, 0.00)	&  0.0235	&  (0.43, 0.00)	\\ 
4294.13	&  Fe \textsc{i} + Ti \textsc{ii}	&  0.1175	&  -0.0040	&  (-0.69, 0.00)	&  -0.0007	&  (0.02, 0.35)	\\ 
4298.99	&  Ca \textsc{i}	&  0.0488	&  -0.0013	&  (-0.13, 0.00)	&  0.0419	&  (0.34, 0.00)	\\ 
4318.65	&  Ca \textsc{i}	&  0.0818	&  -0.0007	&  (-0.05, 0.01)	&  0.0155	&  (0.44, 0.00)	\\ 
4320.74	&  Sc \textsc{ii}	&  0.1094	&  0.0019	&  (0.66, 0.00)	&  0.0053	&  (0.18, 0.00)	\\ 
4325.77	&  Fe \textsc{i}	&  0.1348	&  -0.0034	&  (-0.20, 0.00)	&  0.0059	&  (0.19, 0.00)	\\ 
4330.02	&  V \textsc{i}	&  0.0710	&  -0.0018	&  (-0.74, 0.00)	&  0.0016	&  (0.19, 0.00)	\\ 
4337.05	&  Fe \textsc{i}	&  0.0803	&  -0.0011	&  (-0.38, 0.00)	&  0.0256	&  (0.49, 0.00)	\\ 
4337.92	&  Ti \textsc{ii}	&  0.0860	&  0.0026	&  (0.66, 0.00)	&  0.0024	&  (0.12, 0.00)	\\ 
4340.49	&  H$\gamma$	&  0.3110	&  0.0222	&  (0.94, 0.00)	&  -0.0007	&  (0.22, 0.00)	\\ 
4341.00	&  V \textsc{i}	&  0.0587	&  -0.0014	&  (-0.67, 0.00)	&  0.0133	&  (0.43, 0.00)	\\ 
4351.77	&  Fe \textsc{ii}	&  0.1050	&  -0.0032	&  (-0.67, 0.00)	&  0.0199	&  (0.42, 0.00)	\\ 
4351.91	&  Mg \textsc{i}	&  0.1116	&  -0.0011	&  (-0.32, 0.00)	&  0.0198	&  (0.31, 0.00)	\\ 
4352.74	&  Fe \textsc{i}	&  0.0854	&  -0.0001	&  (0.19, 0.00)	&  0.0031	&  (0.45, 0.00)	\\ 
4352.87	&  V \textsc{i}	&  0.1184	&  -0.0019	&  (-0.70, 0.00)	&  0.0182	&  (0.54, 0.00)	\\ 
4367.91	&  Fe \textsc{i}	&  0.0601	&  -0.0001	&  (0.20, 0.00)	&  0.0126	&  (0.29, 0.00)	\\ 
4368.05	&  V \textsc{i}	&  0.0365	&  -0.0011	&  (-0.50, 0.00)	&  -0.0024	&  (0.13, 0.00)	\\ 
4371.28	&  Cr \textsc{i}	&  0.0861	&  0.0003	&  (0.30, 0.00)	&  0.0178	&  (0.34, 0.00)	\\ 
4374.16	&  Cr \textsc{i}	&  0.0995	&  -0.0017	&  (-0.60, 0.00)	&  0.0097	&  (0.51, 0.00)	\\ 
4374.47	&  Sc \textsc{ii}	&  0.0581	&  -0.0003	&  (-0.08, 0.00)	&  0.0122	&  (0.13, 0.00)	\\ 
4375.94	&  Fe \textsc{i}	&  0.0693	&  -0.0098	&  (-0.95, 0.00)	&  0.0364	&  (0.42, 0.00)	\\ 
4379.23	&  V \textsc{i}	&  0.1136	&  -0.0040	&  (-0.74, 0.00)	&  0.0188	&  (0.41, 0.00)	\\ 
4383.55	&  Fe \textsc{i}	&  1.8531	&  -0.0406	&  (-0.89, 0.00)	&  0.0017	&  (0.31, 0.00)	\\ 
4389.25	&  Fe \textsc{i}	&  0.0636	&  -0.0001	&  (-0.26, 0.00)	&  0.0384	&  (0.54, 0.00)	\\ 
4389.99	&  V \textsc{i}	&  0.1018	&  -0.0014	&  (-0.35, 0.00)	&  0.0126	&  (0.17, 0.00)	\\ 
4395.04	&  Ti \textsc{ii}	&  0.1283	&  0.0025	&  (0.78, 0.00)	&  0.0083	&  (-0.01, 0.52)	\\ 
4395.23	&  V \textsc{i}	&  0.0838	&  -0.0008	&  (-0.60, 0.00)	&  0.0303	&  (0.60, 0.00)	\\ 
4399.77	&  Ti \textsc{ii}	&  0.1057	&  0.0019	&  (0.54, 0.00)	&  -0.0032	&  (0.13, 0.00)	\\ 
4404.76	&  Fe \textsc{i}	&  1.1583	&  -0.0252	&  (-0.90, 0.00)	&  -0.0007	&  (-0.07, 0.00)	\\ 
4406.64	&  V \textsc{i}	&  0.0970	&  -0.0022	&  (-0.81, 0.00)	&  0.0115	&  (0.38, 0.00)	\\ 
4407.68	&  Fe \textsc{i}	&  0.1438	&  -0.0038	&  (-0.76, 0.00)	&  -0.0185	&  (-0.48, 0.00)	\\ 
4408.20	&  V \textsc{i}	&  0.0932	&  -0.0006	&  (-0.41, 0.00)	&  0.0096	&  (0.40, 0.00)	\\ 
4408.47	&  Fe \textsc{i}	&  0.1700	&  -0.0040	&  (-0.67, 0.00)	&  0.0053	&  (0.37, 0.00)	\\ 
4416.47	&  V \textsc{i}	&  0.0816	&  -0.0018	&  (-0.64, 0.00)	&  -0.0007	&  (0.44, 0.00)	\\ 
4421.57	&  V \textsc{i}	&  0.0480	&  -0.0016	&  (-0.66, 0.00)	&  0.0171	&  (0.42, 0.00)	\\ 
4426.02	&  V \textsc{i} + Ti \textsc{i}	&  0.0861	&  -0.0035	&  (-0.76, 0.00)	&  -0.0057	&  (-0.24, 0.00)	\\ 
4427.32	&  Fe \textsc{i}	&  0.0942	&  -0.0093	&  (-0.94, 0.00)	&  0.0200	&  (0.37, 0.00)	\\ 
4428.52	&  V \textsc{i} + Cr \textsc{i}	&  0.0883	&  -0.0019	&  (-0.70, 0.00)	&  0.0090	&  (0.34, 0.00)	\\ 
4429.79	&  V \textsc{i}	&  0.0775	&  -0.0016	&  (-0.77, 0.00)	&  0.0022	&  (0.36, 0.00)	\\ 
4435.15	&  Fe \textsc{i}	&  0.0773	&  -0.0011	&  (-0.58, 0.00)	&  0.0176	&  (0.57, 0.00)	\\ 
4436.14	&  V \textsc{i}	&  0.0581	&  -0.0015	&  (-0.76, 0.00)	&  -0.0001	&  (0.19, 0.00)	\\ 
4437.83	&  V \textsc{i}	&  0.0780	&  -0.0021	&  (-0.67, 0.00)	&  0.0066	&  (0.53, 0.00)	\\ 
4439.88	&  Fe \textsc{i}	&  0.0668	&  0.0018	&  (0.52, 0.00)	&  0.0023	&  (0.16, 0.00)	\\ 
4441.69	&  Ti \textsc{ii} + V \textsc{i}	&  0.0876	&  -0.0019	&  (-0.77, 0.00)	&  0.0060	&  (0.31, 0.00)	\\ 
4442.34	&  Fe \textsc{i}	&  0.0883	&  -0.0007	&  (-0.38, 0.00)	&  0.0287	&  (0.54, 0.00)	\\ 
4443.81	&  Ti \textsc{ii}	&  0.0817	&  0.0001	&  (0.43, 0.00)	&  0.0083	&  (0.18, 0.00)	\\ 
4444.21	&  V \textsc{i}	&  0.0854	&  -0.0015	&  (-0.72, 0.00)	&  0.0174	&  (0.61, 0.00)	\\ 
4451.59	&  Mn \textsc{i}	&  0.1196	&  -0.0005	&  (0.33, 0.00)	&  0.0030	&  (0.18, 0.00)	\\ 
4452.01	&  V \textsc{i}	&  0.0775	&  -0.0014	&  (-0.60, 0.00)	&  0.0014	&  (0.02, 0.39)	\\ 
4454.39	&  Fe \textsc{i}	&  0.1336	&  0.0008	&  (0.53, 0.00)	&  0.0013	&  (0.15, 0.00)	\\ 
4455.32	&  Ti \textsc{i} + Mn \textsc{i}	&  0.1220	&  -0.0014	&  (-0.06, 0.00)	&  0.0035	&  (0.03, 0.10)	\\ 
4455.89	&  Ca \textsc{i}	&  0.0712	&  -0.0012	&  (-0.41, 0.00)	&  0.0301	&  (0.23, 0.00)	\\ 
4457.50	&  Mn \textsc{i} + V \textsc{i}	&  0.1576	&  -0.0030	&  (-0.73, 0.00)	&  -0.0065	&  (-0.22, 0.00)	\\ 
4459.76	&  V \textsc{i}	&  0.0972	&  -0.0024	&  (-0.71, 0.00)	&  -0.0019	&  (0.08, 0.00)	\\ 
4460.30	&  V \textsc{i}	&  0.1294	&  -0.0044	&  (-0.83, 0.00)	&  0.0170	&  (0.36, 0.00)	\\ 
4461.66	&  Fe \textsc{i}	&  0.1144	&  -0.0078	&  (-0.93, 0.00)	&  0.0171	&  (0.64, 0.00)	\\ 
4466.57	&  Fe \textsc{i}	&  0.1262	&  -0.0020	&  (-0.61, 0.00)	&  0.0270	&  (0.51, 0.00)	\\ 
4468.50	&  Ti \textsc{ii}	&  0.1221	&  0.0028	&  (0.80, 0.00)	&  -0.0015	&  (0.03, 0.15)	\\ 

			\bottomrule
		\end{tabular}
	\end{threeparttable}
\end{table*}
\begin{table*}
	\begin{threeparttable}
		\centering
		\contcaption{The activity sensitive lines presented below are also available as a supplementary online table.
The table includes measured center, species, equivalent width of the line (EW, measured from a low activity template),
change in equivalent width ($\Delta$EW, measured from relative spectra), 
Pearson's correlation coefficient and 2-tailed $p$-value for log$S$ vs $\Delta$EW ($\Delta$EW PCC), 
asymmetry of the line -- difference between mean flux in the red and blue wings.
Pearson's correlation coefficient and 2-tailed $p$-value for log$S$ vs Asymmetry (Asym. PCC).
		}
		\label{tab:actlines}
		\begin{tabular}{llp{2cm}p{2cm}p{3cm}p{2cm}p{2cm}}
			\toprule
			Center	& Species	& EW	& $\Delta$EW$^1$	& $\Delta$EW PCC$^2$	& Asym.$^3$	& Asym. PCC$^4$	\\
			\midrule
4472.76	&  Fe \textsc{i}	&  0.1345	&  -0.0018	&  (-0.71, 0.00)	&  0.0016	&  (0.37, 0.00)	\\ 
4482.18	&  Fe \textsc{i}	&  0.1371	&  -0.0085	&  (-0.93, 0.00)	&  0.0688	&  (0.84, 0.00)	\\ 
4482.73	&  Fe \textsc{i}	&  0.1038	&  -0.0007	&  (-0.43, 0.00)	&  -0.0083	&  (-0.45, 0.00)	\\ 
4489.74	&  Fe \textsc{i}	&  0.1263	&  -0.0008	&  (-0.46, 0.00)	&  0.0170	&  (0.70, 0.00)	\\ 
4494.57	&  Fe \textsc{i}	&  0.1555	&  -0.0004	&  (-0.32, 0.00)	&  0.0019	&  (0.30, 0.00)	\\ 
4496.86	&  Cr \textsc{i}	&  0.1475	&  0.0014	&  (0.56, 0.00)	&  0.0054	&  (0.20, 0.00)	\\ 
4501.27	&  Ti \textsc{ii}	&  0.1159	&  0.0020	&  (0.74, 0.00)	&  0.0068	&  (0.20, 0.00)	\\ 
4512.74	&  Ti \textsc{i}	&  0.0959	&  0.0002	&  (0.41, 0.00)	&  0.0045	&  (0.22, 0.00)	\\ 
4514.47	&  Cr \textsc{i}	&  0.1266	&  -0.0011	&  (-0.62, 0.00)	&  0.0004	&  (-0.21, 0.00)	\\ 
4518.03	&  Ti \textsc{i}	&  0.1087	&  0.0010	&  (0.53, 0.00)	&  -0.0018	&  (0.02, 0.33)	\\ 
4522.80	&  Ti \textsc{i}	&  0.1057	&  0.0009	&  (0.53, 0.00)	&  0.0056	&  (0.40, 0.00)	\\ 
4528.62	&  Fe \textsc{i}	&  0.3556	&  -0.0035	&  (-0.73, 0.00)	&  0.0008	&  (0.32, 0.00)	\\ 
4531.15	&  Fe \textsc{i}	&  0.1437	&  -0.0004	&  (-0.37, 0.00)	&  0.0249	&  (0.64, 0.00)	\\ 
4533.25	&  Ti \textsc{i}	&  0.1473	&  0.0012	&  (0.33, 0.00)	&  0.0051	&  (0.20, 0.00)	\\ 
4533.97	&  Ti \textsc{ii}	&  0.1237	&  0.0024	&  (0.79, 0.00)	&  0.0082	&  (0.18, 0.00)	\\ 
4534.78	&  Ti \textsc{i}	&  0.1383	&  0.0014	&  (0.43, 0.00)	&  0.0096	&  (0.39, 0.00)	\\ 
4544.68	&  Ti \textsc{i}	&  0.1524	&  -0.0004	&  (-0.20, 0.00)	&  -0.0071	&  (-0.32, 0.00)	\\ 
4545.35	&  Cr \textsc{i}	&  0.0879	&  -0.0020	&  (-0.66, 0.00)	&  0.0072	&  (0.53, 0.00)	\\ 
4545.96	&  Cr \textsc{i}	&  0.1233	&  0.0013	&  (0.67, 0.00)	&  0.0017	&  (0.23, 0.00)	\\ 
4549.63	&  Ti \textsc{ii}	&  0.2352	&  0.0032	&  (0.79, 0.00)	&  0.0047	&  (0.42, 0.00)	\\ 
4554.03	&  Ba \textsc{ii}	&  0.1755	&  0.0029	&  (0.70, 0.00)	&  -0.0010	&  (-0.00, 0.92)	\\ 
4563.76	&  Ti \textsc{ii}	&  0.1288	&  0.0026	&  (0.75, 0.00)	&  -0.0047	&  (-0.09, 0.00)	\\ 
4571.10	&  Mg \textsc{i}	&  0.1023	&  -0.0046	&  (-0.91, 0.00)	&  0.0130	&  (0.48, 0.00)	\\ 
4571.98	&  Ti \textsc{ii}	&  0.1412	&  0.0042	&  (0.85, 0.00)	&  -0.0014	&  (-0.09, 0.00)	\\ 
4577.18	&  V \textsc{i}	&  0.0753	&  -0.0013	&  (-0.65, 0.00)	&  -0.0019	&  (0.14, 0.00)	\\ 
4586.37	&  V \textsc{i}	&  0.1008	&  -0.0017	&  (-0.67, 0.00)	&  0.0090	&  (0.33, 0.00)	\\ 
4592.66	&  Fe \textsc{i}	&  0.1295	&  0.0006	&  (0.29, 0.00)	&  0.0069	&  (0.19, 0.00)	\\ 
4594.12	&  V \textsc{i}	&  0.1070	&  -0.0024	&  (-0.74, 0.00)	&  0.0084	&  (0.37, 0.00)	\\ 
4600.75	&  Cr \textsc{i}	&  0.1144	&  0.0016	&  (0.53, 0.00)	&  0.0081	&  (0.42, 0.00)	\\ 
4602.95	&  Fe \textsc{i}	&  0.1641	&  -0.0003	&  (0.01, 0.61)	&  0.0044	&  (0.32, 0.00)	\\ 
4616.13	&  Cr \textsc{i}	&  0.1174	&  0.0026	&  (0.80, 0.00)	&  0.0057	&  (0.16, 0.00)	\\ 
4621.93	&  Cr \textsc{i}	&  0.0885	&  -0.0012	&  (-0.70, 0.00)	&  -0.0005	&  (-0.32, 0.00)	\\ 
4626.18	&  Cr \textsc{i}	&  0.1193	&  0.0034	&  (0.77, 0.00)	&  0.0048	&  (0.23, 0.00)	\\ 
4629.35	&  Fe \textsc{ii}	&  0.1070	&  -0.0010	&  (-0.24, 0.00)	&  0.0052	&  (0.36, 0.00)	\\ 
4632.91	&  Fe \textsc{i}	&  0.0648	&  -0.0006	&  (-0.19, 0.00)	&  0.0303	&  (0.44, 0.00)	\\ 
4645.19	&  Ti \textsc{i}	&  0.0542	&  -0.0008	&  (-0.41, 0.00)	&  0.0033	&  (0.29, 0.00)	\\ 
4646.17	&  Cr \textsc{i}	&  0.1376	&  0.0031	&  (0.70, 0.00)	&  0.0039	&  (0.17, 0.00)	\\ 
4647.44	&  Fe \textsc{i}	&  0.1017	&  0.0009	&  (0.58, 0.00)	&  0.0014	&  (0.27, 0.00)	\\ 
4651.29	&  Cr \textsc{i}	&  0.1109	&  0.0019	&  (0.75, 0.00)	&  0.0084	&  (0.44, 0.00)	\\ 
4652.16	&  Cr \textsc{i}	&  0.1411	&  0.0013	&  (0.57, 0.00)	&  0.0081	&  (0.55, 0.00)	\\ 
4656.47	&  Ti \textsc{i}	&  0.0965	&  -0.0002	&  (0.25, 0.00)	&  0.0056	&  (0.41, 0.00)	\\ 
4675.11	&  Fe \textsc{i} + Ti \textsc{i}	&  0.0668	&  -0.0013	&  (-0.56, 0.00)	&  0.0044	&  (0.37, 0.00)	\\ 
4681.91	&  Ti \textsc{i}	&  0.0999	&  0.0003	&  (0.19, 0.00)	&  0.0109	&  (0.57, 0.00)	\\ 
4715.30	&  Ti \textsc{i}	&  0.0382	&  -0.0008	&  (-0.66, 0.00)	&  0.0096	&  (0.28, 0.00)	\\ 
4722.16	&  Zn \textsc{i}	&  0.0763	&  0.0002	&  (0.16, 0.00)	&  -0.0103	&  (-0.32, 0.00)	\\ 
4722.61	&  Ti \textsc{i}	&  0.0513	&  -0.0010	&  (-0.47, 0.00)	&  0.0028	&  (0.16, 0.00)	\\ 
4723.16	&  Ti \textsc{i}	&  0.0915	&  -0.0018	&  (-0.70, 0.00)	&  -0.0017	&  (-0.15, 0.00)	\\ 
4733.60	&  Fe \textsc{i}	&  0.1151	&  0.0014	&  (0.59, 0.00)	&  0.0042	&  (0.09, 0.00)	\\ 
4736.78	&  Fe \textsc{i}	&  0.1301	&  0.0005	&  (0.22, 0.00)	&  0.0019	&  (0.31, 0.00)	\\ 
4754.04	&  Mn \textsc{i}	&  0.1578	&  -0.0007	&  (-0.22, 0.00)	&  0.0083	&  (0.40, 0.00)	\\ 
4759.27	&  Ti \textsc{i}	&  0.0704	&  -0.0002	&  (-0.12, 0.00)	&  0.0056	&  (0.20, 0.00)	\\ 
4761.53	&  Mn \textsc{i}	&  0.0860	&  -0.0006	&  (-0.08, 0.00)	&  0.0046	&  (-0.05, 0.01)	\\ 
4762.38	&  Mn \textsc{i}	&  0.1255	&  -0.0000	&  (-0.02, 0.45)	&  0.0099	&  (0.26, 0.00)	\\ 
4772.82	&  Fe \textsc{i}	&  0.1156	&  0.0004	&  (0.13, 0.00)	&  0.0062	&  (0.50, 0.00)	\\ 
4783.42	&  Mn \textsc{i}	&  0.2302	&  -0.0005	&  (-0.15, 0.00)	&  0.0039	&  (0.33, 0.00)	\\ 
4823.51	&  Mn \textsc{i}	&  0.1629	&  0.0004	&  (0.06, 0.00)	&  0.0117	&  (0.37, 0.00)	\\ 
4827.46	&  V \textsc{i}	&  0.0739	&  -0.0025	&  (-0.89, 0.00)	&  0.0039	&  (0.09, 0.00)	\\ 
4831.65	&  V \textsc{i}	&  0.0713	&  -0.0020	&  (-0.79, 0.00)	&  0.0085	&  (0.42, 0.00)	\\ 
4832.43	&  V \textsc{i}	&  0.0607	&  -0.0016	&  (-0.66, 0.00)	&  0.0065	&  (0.22, 0.00)	\\ 
4840.88	&  Ti \textsc{i} + Fe \textsc{i}	&  0.0952	&  0.0006	&  (0.41, 0.00)	&  0.0019	&  (0.22, 0.00)	\\ 

			\bottomrule
		\end{tabular}
	\end{threeparttable}
\end{table*}
\begin{table*}
	\begin{threeparttable}
		\centering
		\contcaption{The activity sensitive lines presented below are also available as a supplementary online table.
The table includes measured center, species, equivalent width of the line (EW, measured from a low activity template),
change in equivalent width ($\Delta$EW, measured from relative spectra), 
Pearson's correlation coefficient and 2-tailed $p$-value for log$S$ vs $\Delta$EW ($\Delta$EW PCC), 
asymmetry of the line -- difference between mean flux in the red and blue wings.
Pearson's correlation coefficient and 2-tailed $p$-value for log$S$ vs Asymmetry (Asym. PCC).
		}
		\label{tab:actlines}
		\begin{tabular}{llp{2cm}p{2cm}p{3cm}p{2cm}p{2cm}}
			\toprule
			Center	& Species	& EW	& $\Delta$EW$^1$	& $\Delta$EW PCC$^2$	& Asym.$^3$	& Asym. PCC$^4$	\\
			\midrule
4851.50	&  V \textsc{i}	&  0.0784	&  -0.0024	&  (-0.81, 0.00)	&  0.0060	&  (0.20, 0.00)	\\ 
4859.75	&  Fe \textsc{i}	&  0.1710	&  0.0024	&  (0.63, 0.00)	&  0.0061	&  (0.21, 0.00)	\\ 
4861.33	&  H$\beta$	&  0.5264	&  0.0286	&  (0.94, 0.00)	&  -0.0062	&  (-0.68, 0.00)	\\ 
4875.49	&  V \textsc{i}	&  0.0897	&  -0.0019	&  (-0.73, 0.00)	&  -0.0002	&  (0.19, 0.00)	\\ 
4881.56	&  V \textsc{i}	&  0.1124	&  -0.0020	&  (-0.81, 0.00)	&  0.0098	&  (0.40, 0.00)	\\ 
4890.76	&  Fe \textsc{i}	&  0.1111	&  0.0005	&  (0.27, 0.00)	&  0.0083	&  (0.33, 0.00)	\\ 
4891.50	&  Fe \textsc{i}	&  0.0895	&  0.0002	&  (0.02, 0.39)	&  0.0213	&  (0.38, 0.00)	\\ 
4920.51	&  Fe \textsc{i}	&  0.5394	&  -0.0043	&  (-0.63, 0.00)	&  0.0054	&  (0.43, 0.00)	\\ 
4923.93	&  Fe \textsc{ii}	&  0.1256	&  0.0029	&  (0.82, 0.00)	&  0.0045	&  (0.16, 0.00)	\\ 
4934.08	&  Ba \textsc{ii}	&  0.1514	&  -0.0001	&  (0.25, 0.00)	&  -0.0072	&  (-0.54, 0.00)	\\ 
4938.82	&  Fe \textsc{i}	&  0.1426	&  0.0005	&  (0.55, 0.00)	&  0.0047	&  (0.41, 0.00)	\\ 
4939.69	&  Fe \textsc{i}	&  0.1204	&  0.0001	&  (-0.12, 0.00)	&  0.0102	&  (0.60, 0.00)	\\ 
4942.49	&  Cr \textsc{i} + Fe \textsc{i}	&  0.1204	&  -0.0015	&  (-0.37, 0.00)	&  0.0057	&  (0.37, 0.00)	\\ 
4981.74	&  Ti \textsc{i}	&  0.1480	&  0.0014	&  (0.48, 0.00)	&  0.0012	&  (0.24, 0.00)	\\ 
4991.07	&  Ti \textsc{i}	&  0.1501	&  -0.0002	&  (0.48, 0.00)	&  0.0114	&  (0.56, 0.00)	\\ 
4994.13	&  Fe \textsc{i}	&  0.1072	&  -0.0006	&  (-0.54, 0.00)	&  0.0233	&  (0.75, 0.00)	\\ 
4999.51	&  Ti \textsc{i}	&  0.1412	&  0.0005	&  (0.53, 0.00)	&  0.0041	&  (0.21, 0.00)	\\ 
5007.22	&  Ti \textsc{i} + Fe \textsc{i}	&  0.1567	&  -0.0007	&  (-0.36, 0.00)	&  0.0210	&  (0.68, 0.00)	\\ 
5009.65	&  Ti \textsc{i}	&  0.0625	&  -0.0009	&  (-0.57, 0.00)	&  0.0106	&  (0.51, 0.00)	\\ 
5012.08	&  Fe \textsc{i}	&  0.1013	&  -0.0056	&  (-0.93, 0.00)	&  0.0202	&  (0.61, 0.00)	\\ 
5013.30	&  Ti \textsc{i} + Cr \textsc{i}	&  0.0973	&  -0.0004	&  (-0.40, 0.00)	&  0.0055	&  (0.35, 0.00)	\\ 
5014.23	&  Ti \textsc{i} + Ni \textsc{i}	&  0.1650	&  -0.0030	&  (-0.68, 0.00)	&  0.0081	&  (0.18, 0.00)	\\ 
5016.17	&  Ti \textsc{i}	&  0.0759	&  -0.0002	&  (0.18, 0.00)	&  0.0085	&  (0.25, 0.00)	\\ 
5018.44	&  Fe \textsc{ii}	&  0.1515	&  0.0028	&  (0.88, 0.00)	&  -0.0003	&  (-0.09, 0.00)	\\ 
5020.03	&  Ti \textsc{i}	&  0.0629	&  -0.0004	&  (-0.02, 0.28)	&  0.0198	&  (0.50, 0.00)	\\ 
5022.87	&  Ti \textsc{i}	&  0.1026	&  0.0009	&  (0.50, 0.00)	&  0.0061	&  (0.31, 0.00)	\\ 
5024.85	&  Ti \textsc{i}	&  0.1063	&  0.0004	&  (0.47, 0.00)	&  0.0033	&  (0.04, 0.04)	\\ 
5039.96	&  Ti \textsc{i}	&  0.1119	&  0.0015	&  (0.55, 0.00)	&  0.0077	&  (0.33, 0.00)	\\ 
5040.90	&  Fe \textsc{i}	&  0.0863	&  -0.0005	&  (-0.49, 0.00)	&  -0.0038	&  (-0.16, 0.00)	\\ 
5041.08	&  Fe \textsc{i}	&  0.0975	&  -0.0016	&  (-0.68, 0.00)	&  0.0003	&  (0.24, 0.00)	\\ 
5041.76	&  Fe \textsc{i}	&  0.0769	&  -0.0019	&  (-0.69, 0.00)	&  0.0252	&  (0.62, 0.00)	\\ 
5051.64	&  Fe \textsc{i}	&  0.0649	&  -0.0040	&  (-0.88, 0.00)	&  0.0241	&  (0.63, 0.00)	\\ 
5060.08	&  Fe \textsc{i}	&  0.0733	&  -0.0008	&  (-0.55, 0.00)	&  0.0093	&  (0.27, 0.00)	\\ 
5064.66	&  Ti \textsc{i}	&  0.0841	&  -0.0005	&  (-0.32, 0.00)	&  0.0056	&  (0.30, 0.00)	\\ 
5068.77	&  Fe \textsc{i}	&  0.1487	&  -0.0004	&  (0.14, 0.00)	&  0.0074	&  (0.24, 0.00)	\\ 
5079.75	&  Fe \textsc{i}	&  0.1031	&  -0.0018	&  (-0.25, 0.00)	&  0.0117	&  (0.38, 0.00)	\\ 
5083.34	&  Fe \textsc{i}	&  0.1419	&  -0.0006	&  (-0.10, 0.00)	&  0.0104	&  (0.54, 0.00)	\\ 
5098.70	&  Fe \textsc{i}	&  0.0903	&  -0.0003	&  (-0.06, 0.00)	&  0.0133	&  (0.63, 0.00)	\\ 
5105.54	&  Cu \textsc{i}	&  0.1205	&  -0.0007	&  (-0.28, 0.00)	&  0.0020	&  (0.08, 0.00)	\\ 
5107.45	&  Fe \textsc{i}	&  0.1350	&  -0.0014	&  (-0.53, 0.00)	&  0.0102	&  (0.67, 0.00)	\\ 
5107.65	&  Fe \textsc{i}	&  0.1289	&  0.0003	&  (0.43, 0.00)	&  -0.0034	&  (0.32, 0.00)	\\ 
5110.41	&  Fe \textsc{i}	&  0.1019	&  -0.0080	&  (-0.96, 0.00)	&  0.0471	&  (0.84, 0.00)	\\ 
5115.40	&  Ni \textsc{i}	&  0.0886	&  0.0005	&  (0.20, 0.00)	&  -0.0110	&  (-0.54, 0.00)	\\ 
5123.73	&  Fe \textsc{i}	&  0.1333	&  -0.0010	&  (-0.45, 0.00)	&  0.0070	&  (0.49, 0.00)	\\ 
5127.36	&  Fe \textsc{i}	&  0.1189	&  0.0001	&  (0.14, 0.00)	&  0.0155	&  (0.60, 0.00)	\\ 
5142.93	&  Ni \textsc{i}	&  0.0962	&  -0.0014	&  (-0.72, 0.00)	&  0.0306	&  (0.74, 0.00)	\\ 
5147.48	&  Ti \textsc{i}	&  0.0618	&  -0.0010	&  (-0.60, 0.00)	&  0.0182	&  (0.56, 0.00)	\\ 
5150.85	&  Fe \textsc{i}	&  0.1391	&  -0.0017	&  (-0.61, 0.00)	&  0.0175	&  (0.72, 0.00)	\\ 
5151.91	&  Fe \textsc{i}	&  0.1317	&  -0.0004	&  (-0.06, 0.00)	&  0.0012	&  (0.28, 0.00)	\\ 
5152.19	&  Ti \textsc{i}	&  0.0669	&  -0.0009	&  (-0.47, 0.00)	&  0.0095	&  (0.52, 0.00)	\\ 
5166.29	&  Fe \textsc{i}	&  0.0895	&  -0.0042	&  (-0.91, 0.00)	&  0.0177	&  (0.27, 0.00)	\\ 
5167.33	&  Mg \textsc{i}	&  0.1383	&  -0.0050	&  (-0.87, 0.00)	&  0.0174	&  (0.47, 0.00)	\\ 
5167.49	&  Fe \textsc{i}	&  0.1541	&  -0.0036	&  (-0.82, 0.00)	&  0.0165	&  (0.65, 0.00)	\\ 
5168.90	&  Fe \textsc{i}	&  0.1353	&  -0.0025	&  (-0.82, 0.00)	&  0.0205	&  (0.81, 0.00)	\\ 
5171.60	&  Fe \textsc{i}	&  0.1257	&  -0.0025	&  (-0.73, 0.00)	&  0.0288	&  (0.77, 0.00)	\\ 
5172.69	&  Mg \textsc{i}	&  0.1713	&  -0.0046	&  (-0.77, 0.00)	&  0.0144	&  (0.42, 0.00)	\\ 
5183.61	&  Mg \textsc{i}	&  2.2932	&  -0.0430	&  (-0.90, 0.00)	&  -0.0023	&  (-0.35, 0.00)	\\ 
5191.46	&  Fe \textsc{i}	&  0.1522	&  0.0019	&  (0.67, 0.00)	&  0.0059	&  (0.43, 0.00)	\\ 
5192.97	&  Ti \textsc{i}	&  0.0834	&  -0.0005	&  (-0.34, 0.00)	&  0.0243	&  (0.72, 0.00)	\\ 
5194.95	&  Fe \textsc{i}	&  0.1405	&  -0.0013	&  (-0.49, 0.00)	&  0.0086	&  (0.60, 0.00)	\\ 

			\bottomrule
		\end{tabular}
	\end{threeparttable}
\end{table*}
\begin{table*}
	\begin{threeparttable}
		\centering
		\contcaption{The activity sensitive lines presented below are also available as a supplementary online table.
The table includes measured center, species, equivalent width of the line (EW, measured from a low activity template),
change in equivalent width ($\Delta$EW, measured from relative spectra), 
Pearson's correlation coefficient and 2-tailed $p$-value for log$S$ vs $\Delta$EW ($\Delta$EW PCC), 
asymmetry of the line -- difference between mean flux in the red and blue wings.
Pearson's correlation coefficient and 2-tailed $p$-value for log$S$ vs Asymmetry (Asym. PCC).
		}
		\label{tab:actlines}
		\begin{tabular}{llp{2cm}p{2cm}p{3cm}p{2cm}p{2cm}}
			\toprule
			Center	& Species	& EW	& $\Delta$EW$^1$	& $\Delta$EW PCC$^2$	& Asym.$^3$	& Asym. PCC$^4$	\\
			\midrule
5198.71	&  Fe \textsc{i}	&  0.1069	&  0.0002	&  (0.23, 0.00)	&  0.0129	&  (0.59, 0.00)	\\ 
5202.34	&  Fe \textsc{i}	&  0.1853	&  -0.0031	&  (-0.51, 0.00)	&  -0.0097	&  (-0.22, 0.00)	\\ 
5204.53	&  Cr \textsc{i} + Fe \textsc{i}	&  0.2380	&  -0.0062	&  (-0.87, 0.00)	&  0.0429	&  (0.91, 0.00)	\\ 
5206.04	&  Cr \textsc{i}	&  0.3478	&  -0.0013	&  (-0.33, 0.00)	&  0.0073	&  (0.53, 0.00)	\\ 
5208.43	&  Cr \textsc{i}	&  0.1327	&  0.0006	&  (0.22, 0.00)	&  0.0175	&  (0.50, 0.00)	\\ 
5210.39	&  Ti \textsc{i}	&  0.0980	&  -0.0007	&  (-0.40, 0.00)	&  0.0221	&  (0.65, 0.00)	\\ 
5216.28	&  Fe \textsc{i}	&  0.1129	&  -0.0002	&  (-0.27, 0.00)	&  0.0108	&  (0.46, 0.00)	\\ 
5219.70	&  Ti \textsc{i}	&  0.0669	&  -0.0005	&  (-0.49, 0.00)	&  0.0003	&  (0.51, 0.00)	\\ 
5225.53	&  Fe \textsc{i}	&  0.1026	&  0.0018	&  (0.77, 0.00)	&  0.0110	&  (0.60, 0.00)	\\ 
5227.19	&  Fe \textsc{i}	&  0.1644	&  -0.0046	&  (-0.87, 0.00)	&  0.0132	&  (0.50, 0.00)	\\ 
5238.58	&  Ti \textsc{i}	&  0.0590	&  -0.0013	&  (-0.76, 0.00)	&  0.0013	&  (0.08, 0.00)	\\ 
5247.05	&  Fe \textsc{i}	&  0.0997	&  0.0010	&  (0.47, 0.00)	&  0.0066	&  (0.55, 0.00)	\\ 
5247.57	&  Cr \textsc{i}	&  0.1271	&  0.0009	&  (0.53, 0.00)	&  0.0046	&  (0.50, 0.00)	\\ 
5250.21	&  Fe \textsc{i}	&  0.0974	&  0.0007	&  (0.54, 0.00)	&  0.0084	&  (0.55, 0.00)	\\ 
5254.96	&  Fe \textsc{i}	&  0.1180	&  -0.0011	&  (-0.51, 0.00)	&  0.0006	&  (-0.03, 0.16)	\\ 
5264.17	&  Cr \textsc{i}	&  0.1472	&  -0.0023	&  (-0.70, 0.00)	&  0.0297	&  (0.81, 0.00)	\\ 
5269.54	&  Fe \textsc{i}	&  0.2483	&  -0.0087	&  (-0.94, 0.00)	&  0.0150	&  (0.67, 0.00)	\\ 
5270.32	&  Fe \textsc{i}	&  0.2694	&  -0.0068	&  (-0.90, 0.00)	&  0.0039	&  (0.40, 0.00)	\\ 
5298.28	&  Cr \textsc{i}	&  0.0969	&  -0.0008	&  (-0.32, 0.00)	&  0.0318	&  (0.60, 0.00)	\\ 
5341.03	&  Fe \textsc{i} + Mn \textsc{i}	&  0.1799	&  -0.0052	&  (-0.78, 0.00)	&  0.0263	&  (0.49, 0.00)	\\ 
5345.80	&  Cr \textsc{i}	&  0.1107	&  -0.0006	&  (-0.05, 0.01)	&  0.0243	&  (0.57, 0.00)	\\ 
5348.32	&  Cr \textsc{i}	&  0.1450	&  0.0010	&  (0.47, 0.00)	&  0.0085	&  (0.54, 0.00)	\\ 
5371.50	&  Fe \textsc{i}	&  0.0947	&  -0.0041	&  (-0.89, 0.00)	&  0.0063	&  (0.43, 0.00)	\\ 
5394.67	&  Mn \textsc{i}	&  0.1289	&  -0.0044	&  (-0.90, 0.00)	&  0.0112	&  (0.71, 0.00)	\\ 
5397.13	&  Fe \textsc{i}	&  0.1083	&  -0.0053	&  (-0.90, 0.00)	&  0.0336	&  (0.77, 0.00)	\\ 
5405.78	&  Mn \textsc{i}	&  0.1089	&  -0.0037	&  (-0.82, 0.00)	&  0.0273	&  (0.66, 0.00)	\\ 
5407.42	&  Mn \textsc{i} + Fe \textsc{i}	&  0.1545	&  -0.0025	&  (-0.79, 0.00)	&  0.0024	&  (0.24, 0.00)	\\ 
5409.79	&  Cr \textsc{i}	&  0.1831	&  0.0021	&  (0.54, 0.00)	&  0.0038	&  (0.16, 0.00)	\\ 
5420.35	&  Mn \textsc{i}	&  0.1626	&  -0.0036	&  (-0.82, 0.00)	&  0.0012	&  (0.22, 0.00)	\\ 
5426.25	&  Ti \textsc{i}	&  0.0328	&  -0.0007	&  (-0.70, 0.00)	&  0.0038	&  (0.35, 0.00)	\\ 
5429.70	&  Fe \textsc{i}	&  0.0997	&  -0.0050	&  (-0.89, 0.00)	&  0.0264	&  (0.74, 0.00)	\\ 
5432.55	&  Mn \textsc{i}	&  0.1062	&  -0.0043	&  (-0.86, 0.00)	&  0.0092	&  (0.29, 0.00)	\\ 
5434.53	&  Fe \textsc{i}	&  0.1285	&  -0.0044	&  (-0.87, 0.00)	&  0.0236	&  (0.70, 0.00)	\\ 
5446.92	&  Fe \textsc{i}	&  0.1225	&  -0.0054	&  (-0.88, 0.00)	&  0.0156	&  (0.52, 0.00)	\\ 
5455.61	&  Fe \textsc{i}	&  0.1289	&  -0.0035	&  (-0.85, 0.00)	&  0.0316	&  (0.77, 0.00)	\\ 
5460.50	&  Ti \textsc{i}	&  0.0463	&  -0.0015	&  (-0.59, 0.00)	&  0.0022	&  (0.15, 0.00)	\\ 
5470.63	&  Mn \textsc{i}	&  0.1332	&  -0.0024	&  (-0.72, 0.00)	&  0.0024	&  (0.39, 0.00)	\\ 
5476.91	&  Ni \textsc{i}	&  0.2098	&  0.0005	&  (0.36, 0.00)	&  0.0044	&  (0.17, 0.00)	\\ 
5483.36	&  Co \textsc{i}	&  0.0982	&  -0.0022	&  (-0.67, 0.00)	&  0.0018	&  (0.22, 0.00)	\\ 
5497.52	&  Fe \textsc{i}	&  0.0774	&  -0.0028	&  (-0.78, 0.00)	&  0.0228	&  (0.71, 0.00)	\\ 
5501.47	&  Fe \textsc{i}	&  0.1121	&  -0.0014	&  (-0.38, 0.00)	&  0.0226	&  (0.72, 0.00)	\\ 
5506.78	&  Fe \textsc{i}	&  0.1167	&  -0.0011	&  (-0.62, 0.00)	&  0.0263	&  (0.80, 0.00)	\\ 
5516.77	&  Mn \textsc{i}	&  0.1070	&  -0.0016	&  (-0.77, 0.00)	&  0.0010	&  (0.14, 0.00)	\\ 
5528.41	&  Mg \textsc{i}	&  0.1502	&  -0.0002	&  (-0.06, 0.01)	&  0.0081	&  (0.43, 0.00)	\\ 
5537.77	&  Mn \textsc{i}	&  0.1016	&  -0.0027	&  (-0.72, 0.00)	&  -0.0039	&  (-0.11, 0.00)	\\ 
5569.62	&  Fe \textsc{i}	&  0.1612	&  0.0004	&  (0.21, 0.00)	&  0.0074	&  (0.43, 0.00)	\\ 
5581.97	&  Ca \textsc{i}	&  0.1346	&  0.0009	&  (0.30, 0.00)	&  0.0072	&  (0.26, 0.00)	\\ 
5587.86	&  Ni \textsc{i}	&  0.0572	&  -0.0001	&  (-0.23, 0.00)	&  0.0055	&  (0.22, 0.00)	\\ 
5588.76	&  Ca \textsc{i}	&  0.1893	&  0.0014	&  (0.58, 0.00)	&  0.0053	&  (0.30, 0.00)	\\ 
5590.12	&  Ca \textsc{i}	&  0.0948	&  0.0000	&  (0.21, 0.00)	&  0.0044	&  (0.35, 0.00)	\\ 
5598.49	&  Ca \textsc{i}	&  0.0862	&  0.0002	&  (-0.09, 0.00)	&  0.0078	&  (0.26, 0.00)	\\ 
5624.55	&  Fe \textsc{i}	&  0.1133	&  -0.0002	&  (-0.19, 0.00)	&  0.0170	&  (0.70, 0.00)	\\ 
5627.64	&  V \textsc{i}	&  0.0703	&  -0.0011	&  (-0.56, 0.00)	&  0.0011	&  (0.22, 0.00)	\\ 
5670.85	&  V \textsc{i}	&  0.0744	&  -0.0016	&  (-0.81, 0.00)	&  0.0054	&  (0.42, 0.00)	\\ 
5703.58	&  V \textsc{i}	&  0.0882	&  -0.0011	&  (-0.55, 0.00)	&  -0.0023	&  (0.22, 0.00)	\\ 
5707.00	&  V \textsc{i}	&  0.1001	&  -0.0014	&  (-0.70, 0.00)	&  0.0054	&  (0.28, 0.00)	\\ 
5711.09	&  Mg \textsc{i}	&  0.1149	&  -0.0011	&  (-0.28, 0.00)	&  0.0047	&  (0.29, 0.00)	\\ 
5782.13	&  Fe \textsc{i}	&  0.1319	&  -0.0017	&  (-0.63, 0.00)	&  -0.0017	&  (-0.16, 0.00)	\\ 
5853.68	&  Ba \textsc{ii}	&  0.0621	&  0.0009	&  (0.70, 0.00)	&  0.0034	&  (0.10, 0.00)	\\ 
5857.45	&  Ca \textsc{i}	&  0.1152	&  -0.0001	&  (-0.03, 0.10)	&  0.0068	&  (0.45, 0.00)	\\ 

			\bottomrule
		\end{tabular}
	\end{threeparttable}
\end{table*}
\begin{table*}
	\begin{threeparttable}
		\centering
		\contcaption{The activity sensitive lines presented below are also available as a supplementary online table.
The table includes measured center, species, equivalent width of the line (EW, measured from a low activity template),
change in equivalent width ($\Delta$EW, measured from relative spectra), 
Pearson's correlation coefficient and 2-tailed $p$-value for log$S$ vs $\Delta$EW ($\Delta$EW PCC), 
asymmetry of the line -- difference between mean flux in the red and blue wings.
Pearson's correlation coefficient and 2-tailed $p$-value for log$S$ vs Asymmetry (Asym. PCC).
		}
		\label{tab:actlines}
		\begin{tabular}{llp{2cm}p{2cm}p{3cm}p{2cm}p{2cm}}
			\toprule
			Center	& Species	& EW	& $\Delta$EW$^1$	& $\Delta$EW PCC$^2$	& Asym.$^3$	& Asym. PCC$^4$	\\
			\midrule
5889.96	&  Na \textsc{i}	&  0.1743	&  -0.0046	&  (-0.59, 0.00)	&  -0.0112	&  (-0.13, 0.00)	\\ 
5895.93	&  Na \textsc{i}	&  0.2286	&  -0.0071	&  (-0.69, 0.00)	&  0.0065	&  (0.15, 0.00)	\\ 
6013.50	&  Mn \textsc{i}	&  0.1152	&  -0.0020	&  (-0.61, 0.00)	&  0.0100	&  (0.26, 0.00)	\\ 
6016.64	&  Fe \textsc{i} + Mn \textsc{i}	&  0.1450	&  -0.0026	&  (-0.56, 0.00)	&  -0.0016	&  (-0.07, 0.00)	\\ 
6021.80	&  Mn \textsc{i}	&  0.1176	&  -0.0000	&  (-0.18, 0.00)	&  0.0029	&  (0.27, 0.00)	\\ 
6065.49	&  Fe \textsc{i}	&  0.0951	&  -0.0002	&  (-0.08, 0.00)	&  0.0110	&  (0.56, 0.00)	\\ 
6081.45	&  V \textsc{i}	&  0.0596	&  -0.0013	&  (-0.53, 0.00)	&  0.0047	&  (0.23, 0.00)	\\ 
6082.71	&  Fe \textsc{i}	&  0.0544	&  -0.0002	&  (-0.04, 0.04)	&  0.0124	&  (0.57, 0.00)	\\ 
6085.25	&  Fe \textsc{i} + Ti \textsc{i}	&  0.0626	&  -0.0012	&  (-0.63, 0.00)	&  0.0046	&  (0.16, 0.00)	\\ 
6090.21	&  V \textsc{i}	&  0.0699	&  -0.0013	&  (-0.59, 0.00)	&  0.0077	&  (0.52, 0.00)	\\ 
6102.72	&  Ca \textsc{i}	&  0.2208	&  0.0014	&  (0.53, 0.00)	&  0.0003	&  (0.30, 0.00)	\\ 
6111.65	&  V \textsc{i}	&  0.0568	&  -0.0013	&  (-0.70, 0.00)	&  0.0030	&  (0.29, 0.00)	\\ 
6122.22	&  Ca \textsc{i}	&  0.1828	&  0.0004	&  (0.56, 0.00)	&  0.0078	&  (0.58, 0.00)	\\ 
6136.62	&  Fe \textsc{i}	&  0.1101	&  -0.0010	&  (-0.33, 0.00)	&  0.0120	&  (0.63, 0.00)	\\ 
6141.73	&  Ba \textsc{ii} + Fe \textsc{i}	&  0.0576	&  0.0012	&  (0.84, 0.00)	&  0.0003	&  (0.38, 0.00)	\\ 
6150.15	&  V \textsc{i}	&  0.0579	&  -0.0014	&  (-0.61, 0.00)	&  0.0028	&  (0.36, 0.00)	\\ 
6173.34	&  Fe \textsc{i}	&  0.0964	&  0.0006	&  (0.48, 0.00)	&  0.0004	&  (0.23, 0.00)	\\ 
6191.57	&  Fe \textsc{i}	&  0.1161	&  -0.0008	&  (-0.40, 0.00)	&  0.0081	&  (0.56, 0.00)	\\ 
6199.19	&  V \textsc{i} + Fe \textsc{ii}	&  0.0752	&  -0.0015	&  (-0.66, 0.00)	&  0.0028	&  (0.33, 0.00)	\\ 
6200.32	&  Fe \textsc{i}	&  0.1002	&  0.0009	&  (0.48, 0.00)	&  -0.0038	&  (-0.12, 0.00)	\\ 
6213.44	&  Fe \textsc{i}	&  0.1148	&  0.0015	&  (0.71, 0.00)	&  -0.0038	&  (0.24, 0.00)	\\ 
6213.87	&  V \textsc{i}	&  0.0405	&  -0.0010	&  (-0.44, 0.00)	&  0.0065	&  (0.12, 0.00)	\\ 
6216.36	&  Fe \textsc{i} + V \textsc{i}	&  0.0927	&  -0.0018	&  (-0.66, 0.00)	&  0.0074	&  (0.37, 0.00)	\\ 
6219.29	&  Fe \textsc{i}	&  0.1343	&  0.0014	&  (0.54, 0.00)	&  0.0009	&  (0.08, 0.00)	\\ 
6230.73	&  Fe \textsc{i}	&  0.2055	&  -0.0015	&  (-0.44, 0.00)	&  0.0096	&  (0.76, 0.00)	\\ 
6242.83	&  V \textsc{i}	&  0.0566	&  -0.0011	&  (-0.56, 0.00)	&  0.0015	&  (0.17, 0.00)	\\ 
6243.11	&  V \textsc{i}	&  0.0936	&  -0.0012	&  (-0.70, 0.00)	&  0.0035	&  (0.32, 0.00)	\\ 
6246.32	&  Fe \textsc{i}	&  0.1655	&  0.0000	&  (0.32, 0.00)	&  0.0062	&  (0.57, 0.00)	\\ 
6251.83	&  V \textsc{i}	&  0.0736	&  -0.0013	&  (-0.67, 0.00)	&  0.0013	&  (0.19, 0.00)	\\ 
6252.56	&  Fe \textsc{i}	&  0.1160	&  -0.0002	&  (-0.24, 0.00)	&  0.0177	&  (0.72, 0.00)	\\ 
6254.26	&  Fe \textsc{i}	&  0.1452	&  -0.0002	&  (0.13, 0.00)	&  0.0022	&  (-0.09, 0.00)	\\ 
6256.36	&  Fe \textsc{i} + Ni \textsc{i}	&  0.1238	&  0.0000	&  (0.24, 0.00)	&  0.0037	&  (0.15, 0.00)	\\ 
6265.14	&  Fe \textsc{i}	&  0.1207	&  0.0008	&  (0.43, 0.00)	&  0.0033	&  (0.26, 0.00)	\\ 
6274.66	&  V \textsc{i}	&  0.0495	&  -0.0016	&  (-0.68, 0.00)	&  0.0029	&  (0.18, 0.00)	\\ 
6318.03	&  Fe \textsc{i} + Ca \textsc{i}	&  0.1272	&  -0.0023	&  (-0.47, 0.00)	&  0.0003	&  (-0.19, 0.00)	\\ 
6335.34	&  Fe \textsc{i}	&  0.1055	&  0.0000	&  (0.18, 0.00)	&  0.0014	&  (0.02, 0.26)	\\ 
6358.69	&  Fe \textsc{i}	&  0.1259	&  -0.0021	&  (-0.28, 0.00)	&  -0.0040	&  (-0.29, 0.00)	\\ 
6393.61	&  Fe \textsc{i}	&  0.1250	&  -0.0009	&  (-0.29, 0.00)	&  0.0118	&  (0.61, 0.00)	\\ 
6400.32	&  Fe \textsc{i}	&  0.0820	&  -0.0004	&  (-0.34, 0.00)	&  0.0053	&  (0.42, 0.00)	\\ 
6408.02	&  Fe \textsc{i}	&  0.1141	&  0.0006	&  (0.48, 0.00)	&  0.0060	&  (0.33, 0.00)	\\ 
6421.36	&  Fe \textsc{i}	&  0.0891	&  -0.0004	&  (-0.22, 0.00)	&  0.0073	&  (0.47, 0.00)	\\ 
6430.85	&  Fe \textsc{i}	&  0.1247	&  -0.0006	&  (-0.00, 1.00)	&  0.0034	&  (0.52, 0.00)	\\ 
6450.20	&  Si \textsc{i}	&  0.1532	&  -0.0034	&  (-0.73, 0.00)	&  0.0056	&  (0.51, 0.00)	\\ 
6562.80	&  H$\alpha$	&  0.7313	&  -0.0151	&  (-0.86, 0.00)	&  0.0027	&  (0.34, 0.00)	\\ 
6643.64	&  Ni \textsc{i}	&  0.0992	&  -0.0015	&  (-0.48, 0.00)	&  -0.0001	&  (-0.02, 0.38)	\\ 

			\bottomrule
		\end{tabular}

	\end{threeparttable}
\end{table*}

%% file: paper.bbl
\begin{thebibliography}{}
\makeatletter
\relax
\def\mn@urlcharsother{\let\do\@makeother \do\$\do\&\do\#\do\^\do\_\do\%\do\~}
\def\mn@doi{\begingroup\mn@urlcharsother \@ifnextchar [ {\mn@doi@}
  {\mn@doi@[]}}
\def\mn@doi@[#1]#2{\def\@tempa{#1}\ifx\@tempa\@empty \href
  {http://dx.doi.org/#2} {doi:#2}\else \href {http://dx.doi.org/#2} {#1}\fi
  \endgroup}
\def\mn@eprint#1#2{\mn@eprint@#1:#2::\@nil}
\def\mn@eprint@arXiv#1{\href {http://arxiv.org/abs/#1} {{\tt arXiv:#1}}}
\def\mn@eprint@dblp#1{\href {http://dblp.uni-trier.de/rec/bibtex/#1.xml}
  {dblp:#1}}
\def\mn@eprint@#1:#2:#3:#4\@nil{\def\@tempa {#1}\def\@tempb {#2}\def\@tempc
  {#3}\ifx \@tempc \@empty \let \@tempc \@tempb \let \@tempb \@tempa \fi \ifx
  \@tempb \@empty \def\@tempb {arXiv}\fi \@ifundefined
  {mn@eprint@\@tempb}{\@tempb:\@tempc}{\expandafter \expandafter \csname
  mn@eprint@\@tempb\endcsname \expandafter{\@tempc}}}

\bibitem[\protect\citeauthoryear{{Anglada-Escud{\'e}} \&
  {Butler}}{{Anglada-Escud{\'e}} \& {Butler}}{2012}]{2012ApJS..200...15A}
{Anglada-Escud{\'e}} G.,  {Butler} R.~P.,  2012, \mn@doi [\apjs]
  {10.1088/0067-0049/200/2/15}, \href
  {http://adsabs.harvard.edu/abs/2012ApJS..200...15A} {200, 15}

\bibitem[\protect\citeauthoryear{{Artigau} et~al.,}{{Artigau}
  et~al.}{2014}]{2014SPIE.9149E..05A}
{Artigau} {\'E}.,  et~al., 2014, in Observatory Operations: Strategies,
  Processes, and Systems V. p. 914905 (\mn@eprint {arXiv} {1406.6927}),
  \mn@doi{10.1117/12.2056385}

\bibitem[\protect\citeauthoryear{{Astropy Collaboration}}{{Astropy
  Collaboration}}{2013}]{2013A&A...558A..33A}
{Astropy Collaboration} 2013, \mn@doi [\aap] {10.1051/0004-6361/201322068},
  \href {https://ui.adsabs.harvard.edu/#abs/2013A&A...558A..33A} {558, A33}

\bibitem[\protect\citeauthoryear{{Berdi{\~n}as}, {Amado}, {Anglada-Escud{\'e}},
  {Rodr{\'{\i}}guez-L{\'o}pez}  \& {Barnes}}{{Berdi{\~n}as}
  et~al.}{2016}]{2016MNRAS.459.3551B}
{Berdi{\~n}as} Z.~M.,  {Amado} P.~J.,  {Anglada-Escud{\'e}} G.,
  {Rodr{\'{\i}}guez-L{\'o}pez} C.,   {Barnes} J.,  2016, \mn@doi [\mnras]
  {10.1093/mnras/stw906}, \href
  {http://adsabs.harvard.edu/abs/2016MNRAS.459.3551B} {459, 3551}

\bibitem[\protect\citeauthoryear{{Berdi{\~n}as}, {Rodr{\'{\i}}guez-L{\'o}pez},
  {Amado}, {Anglada-Escud{\'e}}, {Barnes}, {MacDonald}, {Zechmeister}  \&
  {Sarmiento}}{{Berdi{\~n}as} et~al.}{2017}]{2017MNRAS.469.4268B}
{Berdi{\~n}as} Z.~M.,  {Rodr{\'{\i}}guez-L{\'o}pez} C.,  {Amado} P.~J.,
  {Anglada-Escud{\'e}} G.,  {Barnes} J.~R.,  {MacDonald} J.,  {Zechmeister} M.,
    {Sarmiento} L.~F.,  2017, \mn@doi [\mnras] {10.1093/mnras/stx1140}, \href
  {http://adsabs.harvard.edu/abs/2017MNRAS.469.4268B} {469, 4268}

\bibitem[\protect\citeauthoryear{{Bergmann}, {Endl}, {Hearnshaw}, {Wittenmyer}
  \& {Wright}}{{Bergmann} et~al.}{2015}]{2015IJAsB..14..173B}
{Bergmann} C.,  {Endl} M.,  {Hearnshaw} J.~B.,  {Wittenmyer} R.~A.,   {Wright}
  D.~J.,  2015, \mn@doi [International Journal of Astrobiology]
  {10.1017/S1473550414000317}, \href
  {http://adsabs.harvard.edu/abs/2015IJAsB..14..173B} {14, 173}

\bibitem[\protect\citeauthoryear{{Boisse}, {Bouchy}, {H{\'e}brard}, {Bonfils},
  {Santos}  \& {Vauclair}}{{Boisse} et~al.}{2011}]{2011IAUS..273..281B}
{Boisse} I.,  {Bouchy} F.,  {H{\'e}brard} G.,  {Bonfils} X.,  {Santos} N.,
  {Vauclair} S.,  2011, in {Prasad Choudhary} D.,  {Strassmeier} K.~G.,  eds,
  IAU Symposium Vol. 273, Physics of Sun and Star Spots. pp 281--285
  (\mn@eprint {arXiv} {1012.1452}), \mn@doi{10.1017/S1743921311015389}

\bibitem[\protect\citeauthoryear{{Butler}, {Marcy}, {Williams}, {McCarthy},
  {Dosanjh}  \& {Vogt}}{{Butler} et~al.}{1996}]{1996PASP..108..500B}
{Butler} R.~P.,  {Marcy} G.~W.,  {Williams} E.,  {McCarthy} C.,  {Dosanjh} P.,
   {Vogt} S.~S.,  1996, \mn@doi [\pasp] {10.1086/133755}, \href
  {http://adsabs.harvard.edu/abs/1996PASP..108..500B} {108, 500}

\bibitem[\protect\citeauthoryear{{Cunha}, {Santos}, {Figueira}, {Santerne},
  {Bertaux}  \& {Lovis}}{{Cunha} et~al.}{2014}]{2014A&A...568A..35C}
{Cunha} D.,  {Santos} N.~C.,  {Figueira} P.,  {Santerne} A.,  {Bertaux} J.~L.,
   {Lovis} C.,  2014, \mn@doi [\aap] {10.1051/0004-6361/201423723}, \href
  {http://adsabs.harvard.edu/abs/2014A%26A...568A..35C} {568, A35}

\bibitem[\protect\citeauthoryear{{Davis}, {Cisewski}, {Dumusque}, {Fischer}  \&
  {Ford}}{{Davis} et~al.}{2017}]{2017ApJ...846...59D}
{Davis} A.~B.,  {Cisewski} J.,  {Dumusque} X.,  {Fischer} D.~A.,   {Ford}
  E.~B.,  2017, \mn@doi [\apj] {10.3847/1538-4357/aa8303}, \href
  {http://adsabs.harvard.edu/abs/2017ApJ...846...59D} {846, 59}

\bibitem[\protect\citeauthoryear{{Dumusque}}{{Dumusque}}{2018}]{2018arXiv180901548D}
{Dumusque} X.,  2018, preprint, \href
  {http://adsabs.harvard.edu/abs/2018arXiv180901548D} {} (\mn@eprint {arXiv}
  {1809.01548})

\bibitem[\protect\citeauthoryear{{Dumusque} et~al.,}{{Dumusque}
  et~al.}{2012}]{2012Natur.491..207D}
{Dumusque} X.,  et~al., 2012, \mn@doi [\nat] {10.1038/nature11572}, \href
  {http://adsabs.harvard.edu/abs/2012Natur.491..207D} {491, 207}

\bibitem[\protect\citeauthoryear{{Duncan} et~al.,}{{Duncan}
  et~al.}{1991}]{1991ApJS...76..383D}
{Duncan} D.~K.,  et~al., 1991, \mn@doi [ApJ] {10.1086/191572}, \href
  {http://adsabs.harvard.edu/abs/1991ApJS...76..383D} {76, 383}

\bibitem[\protect\citeauthoryear{{Feng}, {Tuomi}, {Jones}, {Barnes},
  {Anglada-Escud{\'e}}, {Vogt}  \& {Butler}}{{Feng}
  et~al.}{2017}]{2017AJ....154..135F}
{Feng} F.,  {Tuomi} M.,  {Jones} H.~R.~A.,  {Barnes} J.,  {Anglada-Escud{\'e}}
  G.,  {Vogt} S.~S.,   {Butler} R.~P.,  2017, \mn@doi [\aj]
  {10.3847/1538-3881/aa83b4}, \href
  {http://adsabs.harvard.edu/abs/2017AJ....154..135F} {154, 135}

\bibitem[\protect\citeauthoryear{{Gomes da Silva}, {Santos}, {Bonfils},
  {Delfosse}, {Forveille}  \& {Udry}}{{Gomes da Silva}
  et~al.}{2011}]{2011AA...534A..30G}
{Gomes da Silva} J.,  {Santos} N.~C.,  {Bonfils} X.,  {Delfosse} X.,
  {Forveille} T.,   {Udry} S.,  2011, \mn@doi [A\&A]
  {10.1051/0004-6361/201116971}, \href
  {http://adsabs.harvard.edu/abs/2011A%26A...534A..30G} {534, A30}

\bibitem[\protect\citeauthoryear{{Hatzes}}{{Hatzes}}{2013}]{2013ApJ...770..133H}
{Hatzes} A.~P.,  2013, \mn@doi [\apj] {10.1088/0004-637X/770/2/133}, \href
  {http://adsabs.harvard.edu/abs/2013ApJ...770..133H} {770, 133}

\bibitem[\protect\citeauthoryear{{Henry} \& {Newsom}}{{Henry} \&
  {Newsom}}{1996}]{1996PASP..108..242H}
{Henry} G.~W.,  {Newsom} M.~S.,  1996, \mn@doi [\pasp] {10.1086/133716}, \href
  {http://adsabs.harvard.edu/abs/1996PASP..108..242H} {108, 242}

\bibitem[\protect\citeauthoryear{{Hu{\'e}lamo} et~al.,}{{Hu{\'e}lamo}
  et~al.}{2008}]{2008A&A...489L...9H}
{Hu{\'e}lamo} N.,  et~al., 2008, \mn@doi [\aap] {10.1051/0004-6361:200810596},
  \href {http://adsabs.harvard.edu/abs/2008A%26A...489L...9H} {489, L9}

\bibitem[\protect\citeauthoryear{Hunter}{Hunter}{2007}]{Hunter:2007}
Hunter J.~D.,  2007, Computing In Science \& Engineering, 9, 90

\bibitem[\protect\citeauthoryear{{Hussain}}{{Hussain}}{1984}]{1984SoEn...33..217H}
{Hussain} M.,  1984, \mn@doi [Solar Energy] {10.1016/0038-092X(84)90240-8},
  \href {http://adsabs.harvard.edu/abs/1984SoEn...33..217H} {33, 217}

\bibitem[\protect\citeauthoryear{Jones, Oliphant, Peterson  \& Others}{Jones
  et~al.}{2001}]{jones_scipy_2001}
Jones E.,  Oliphant T.,  Peterson P.,   Others 2001, {SciPy}: Open source
  scientific tools for Python, \url {http://www.scipy.org/}

\bibitem[\protect\citeauthoryear{{K{\"u}rster} et~al.,}{{K{\"u}rster}
  et~al.}{2003}]{2003A&A...403.1077K}
{K{\"u}rster} M.,  et~al., 2003, \mn@doi [\aap] {10.1051/0004-6361:20030396},
  \href {http://adsabs.harvard.edu/abs/2003A%26A...403.1077K} {403, 1077}

\bibitem[\protect\citeauthoryear{{Lindegren} \& {Dravins}}{{Lindegren} \&
  {Dravins}}{2003}]{2003AA...401.1185L}
{Lindegren} L.,  {Dravins} D.,  2003, \mn@doi [\aap]
  {10.1051/0004-6361:20030181}, \href
  {http://adsabs.harvard.edu/abs/2003A%26A...401.1185L} {401, 1185}

\bibitem[\protect\citeauthoryear{{Lobel}}{{Lobel}}{2006}]{2006IAUJD...4E..22L}
{Lobel} A.,  2006, in IAU Joint Discussion. p.~22

\bibitem[\protect\citeauthoryear{{Lobel}}{{Lobel}}{2008}]{2008JPhCS.130a2015L}
{Lobel} A.,  2008, in Journal of Physics Conference Series. p. 012015
  (\mn@eprint {arXiv} {0712.1185}), \mn@doi{10.1088/1742-6596/130/1/012015}

\bibitem[\protect\citeauthoryear{{Lobel}}{{Lobel}}{2011}]{2011CaJPh..89..395L}
{Lobel} A.,  2011, \mn@doi [Canadian Journal of Physics] {10.1139/p11-030},
  \href {http://adsabs.harvard.edu/abs/2011CaJPh..89..395L} {89, 395}

\bibitem[\protect\citeauthoryear{{Mayor} \& {Queloz}}{{Mayor} \&
  {Queloz}}{1995}]{1995Natur.378..355M}
{Mayor} M.,  {Queloz} D.,  1995, \mn@doi [\nat] {10.1038/378355a0}, \href
  {http://adsabs.harvard.edu/abs/1995Natur.378..355M} {378, 355}

\bibitem[\protect\citeauthoryear{{Mayor} et~al.,}{{Mayor}
  et~al.}{2003}]{2003Msngr.114...20M}
{Mayor} M.,  et~al., 2003, The Messenger, \href
  {http://adsabs.harvard.edu/abs/2003Msngr.114...20M} {114, 20}

\bibitem[\protect\citeauthoryear{McKinney}{McKinney}{2010}]{mckinney}
McKinney W.,  2010, in van~der Walt S.,  Millman J.,  eds, Proceedings of the
  9th Python in Science Conference. pp 51 -- 56

\bibitem[\protect\citeauthoryear{{O'Toole}, {Tinney}  \& {Jones}}{{O'Toole}
  et~al.}{2008}]{2008MNRAS.386..516O}
{O'Toole} S.~J.,  {Tinney} C.~G.,   {Jones} H.~R.~A.,  2008, \mn@doi [\mnras]
  {10.1111/j.1365-2966.2008.13061.x}, \href
  {http://adsabs.harvard.edu/abs/2008MNRAS.386..516O} {386, 516}

\bibitem[\protect\citeauthoryear{{Queloz} et~al.,}{{Queloz}
  et~al.}{2001}]{2001A&A...379..279Q}
{Queloz} D.,  et~al., 2001, \mn@doi [\aap] {10.1051/0004-6361:20011308}, \href
  {http://adsabs.harvard.edu/abs/2001A%26A...379..279Q} {379, 279}

\bibitem[\protect\citeauthoryear{{Rajpaul}, {Aigrain}  \& {Roberts}}{{Rajpaul}
  et~al.}{2016}]{2016MNRAS.456L...6R}
{Rajpaul} V.,  {Aigrain} S.,   {Roberts} S.,  2016, \mn@doi [\mnras]
  {10.1093/mnrasl/slv164}, \href
  {http://adsabs.harvard.edu/abs/2016MNRAS.456L...6R} {456, L6}

\bibitem[\protect\citeauthoryear{{Saar} \& {Donahue}}{{Saar} \&
  {Donahue}}{1997}]{1997ApJ...485..319S}
{Saar} S.~H.,  {Donahue} R.~A.,  1997, \mn@doi [\apj] {10.1086/304392}, \href
  {http://adsabs.harvard.edu/abs/1997ApJ...485..319S} {485, 319}

\bibitem[\protect\citeauthoryear{{Santos}, {Mayor}, {Naef}, {Queloz}  \&
  {Udry}}{{Santos} et~al.}{2001}]{2001astro.ph..1377S}
{Santos} N.~C.,  {Mayor} M.,  {Naef} D.,  {Queloz} D.,   {Udry} S.,  2001,
  ArXiv Astrophysics e-prints, \href
  {http://adsabs.harvard.edu/abs/2001astro.ph..1377S} {}

\bibitem[\protect\citeauthoryear{{Santos}, {Gomes da Silva}, {Lovis}  \&
  {Melo}}{{Santos} et~al.}{2010}]{2010A&A...511A..54S}
{Santos} N.~C.,  {Gomes da Silva} J.,  {Lovis} C.,   {Melo} C.,  2010, \mn@doi
  [\aap] {10.1051/0004-6361/200913433}, \href
  {http://adsabs.harvard.edu/abs/2010A%26A...511A..54S} {511, A54}

\bibitem[\protect\citeauthoryear{{Struve}}{{Struve}}{1952}]{1952Obs....72..199S}
{Struve} O.,  1952, The Observatory, \href
  {http://adsabs.harvard.edu/abs/1952Obs....72..199S} {72, 199}

\bibitem[\protect\citeauthoryear{{Thompson}, {Watson}, {de Mooij}  \&
  {Jess}}{{Thompson} et~al.}{2017}]{2017MNRAS.468L..16T}
{Thompson} A.~P.~G.,  {Watson} C.~A.,  {de Mooij} E.~J.~W.,   {Jess} D.~B.,
  2017, \mn@doi [\mnras] {10.1093/mnrasl/slx018}, \href
  {http://adsabs.harvard.edu/abs/2017MNRAS.468L..16T} {468, L16}

\bibitem[\protect\citeauthoryear{Van Der~Walt, Colbert  \& Varoquaux}{Van
  Der~Walt et~al.}{2011}]{van2011numpy}
Van Der~Walt S.,  Colbert S.~C.,   Varoquaux G.,  2011, Computing in Science \&
  Engineering, 13, 22

\bibitem[\protect\citeauthoryear{{Wilson}}{{Wilson}}{1978}]{1978ApJ...226..379W}
{Wilson} O.~C.,  1978, \mn@doi [\apj] {10.1086/156618}, \href
  {http://adsabs.harvard.edu/abs/1978ApJ...226..379W} {226, 379}

\bibitem[\protect\citeauthoryear{{Wise}, {Dodson-Robinson}, {Bevenour}  \&
  {Provini}}{{Wise} et~al.}{2018}]{2018AJ....156..180W}
{Wise} A.~W.,  {Dodson-Robinson} S.~E.,  {Bevenour} K.,   {Provini} A.,  2018,
  \mn@doi [\aj] {10.3847/1538-3881/aadd94}, \href
  {http://adsabs.harvard.edu/abs/2018AJ....156..180W} {156, 180}

\bibitem[\protect\citeauthoryear{{Wolszczan} \& {Frail}}{{Wolszczan} \&
  {Frail}}{1992}]{1992Natur.355..145W}
{Wolszczan} A.,  {Frail} D.~A.,  1992, \mn@doi [\nat] {10.1038/355145a0}, \href
  {http://adsabs.harvard.edu/abs/1992Natur.355..145W} {355, 145}

\bibitem[\protect\citeauthoryear{{Wright}, {Wittenmyer}, {Tinney}, {Bentley}
  \& {Zhao}}{{Wright} et~al.}{2016}]{2016ApJ...817L..20W}
{Wright} D.~J.,  {Wittenmyer} R.~A.,  {Tinney} C.~G.,  {Bentley} J.~S.,
  {Zhao} J.,  2016, \mn@doi [\apjl] {10.3847/2041-8205/817/2/L20}, \href
  {http://adsabs.harvard.edu/abs/2016ApJ...817L..20W} {817, L20}

\bibitem[\protect\citeauthoryear{{Zhao}, {Fischer}, {Brewer}, {Giguere}  \&
  {Rojas-Ayala}}{{Zhao} et~al.}{2018}]{2018AJ....155...24Z}
{Zhao} L.,  {Fischer} D.~A.,  {Brewer} J.,  {Giguere} M.,   {Rojas-Ayala} B.,
  2018, \mn@doi [\aj] {10.3847/1538-3881/aa9bea}, \href
  {http://adsabs.harvard.edu/abs/2018AJ....155...24Z} {155, 24}

\makeatother
\end{thebibliography}
